\documentclass[a4paper,11pt]{article}
\pdfoutput=1 

\usepackage{jcappub} 

\usepackage[T1]{fontenc} 
\usepackage{multirow}
\usepackage{natbib}
\usepackage{aas_macros}
\usepackage{amsmath}
\usepackage{color}
\usepackage{enumerate}
\usepackage{multirow}
\usepackage{bbm}
\usepackage{hyperref}
\usepackage{tabularx}
\usepackage{xfrac}
\usepackage{arydshln}
\usepackage{mathrsfs}
\usepackage{enumitem}
\usepackage{subfigure}
\usepackage{pdflscape}
\usepackage{bbold}
\usepackage{bm}

\usepackage{caption}
\usepackage{newfloat}

\DeclareFloatingEnvironment[name={List}]{enumcnt}

\definecolor{internationalkleinblue}{rgb}{0.0, 0.18, 0.65}

\hypersetup{urlcolor=internationalkleinblue, linkcolor=internationalkleinblue, citecolor=internationalkleinblue}

\title{Consistency tests in cosmology using relative entropy}

\author[a, b]{Andrina Nicola,}
\author[a]{Adam Amara,}
\author[a]{Alexandre Refregier}

\affiliation[a]{Department of Physics, ETH Z\"urich, Wolfgang-Pauli-Strasse 27, CH-8093 Z\"urich, Switzerland}
\affiliation[b]{Department of Astrophysical Sciences, Princeton University, Princeton, NJ 08544, USA}

\emailAdd{anicola@astro.princeton.edu, adam.amara@phys.ethz.ch, alexandre.refregier@phys.ethz.ch}

\abstract{With the high-precision data from current and upcoming experiments, it becomes increasingly important to perform consistency tests of the standard cosmological model. In this work, we focus on consistency measures between different data sets and methods that allow us to assess the goodness of fit of different models. We address both of these questions using the relative entropy or Kullback-Leibler (KL) divergence \cite{Kullback:1951}. First, we revisit the relative entropy as a consistency measure between data sets and further investigate some of its key properties, such as asymmetry and path dependence. We then introduce a novel model rejection framework, which is based on the relative entropy and the posterior predictive distribution. We validate the method on several toy models and apply it to Type Ia supernovae data from the JLA and CMB constraints from Planck 2015, testing the consistency of the data with six different cosmological models.}

\begin{document}
\maketitle
\flushbottom

\section{Introduction}

The $\Lambda$ Cold Dark Matter ($\Lambda$CDM) cosmological model has been remarkably successful in describing cosmological data from many different probes across a wide range of length scales and has thus passed the tests it has been subject to so far. Despite its success, several open questions remain, such as the nature of dark matter and dark energy and the origin of primordial fluctuations. Current and future cosmological experiments, such as the Dark Energy Survey (DES\footnote{\tt{https://www.darkenergysurvey.org.}}), the Kilo Degree Survey (KiDS\footnote{\tt{http://kids.strw.leidenuniv.nl.}}), The Hyper Suprime Cam (HSC\footnote{\tt{http://hsc.mtk.nao.ac.jp/ssp.}}), Planck\footnote{\tt{https://www.cosmos.esa.int/web/planck.}}, the Dark Energy Spectroscopic Instrument (DESI\footnote{\tt{http://desi.lbl.gov}.}), the Prime Focus Spectrograph (PFS\footnote{\tt{http://pfs.ipmu.jp}.}), the Large Synoptic Survey Telescope (LSST\footnote{\tt{http://www.lsst.org}.}), Euclid\footnote{\tt{http://sci.esa.int/euclid}.}, the Wide Field Infrared Survey Telescope (WFIRST\footnote{\tt{https://wfirst.gsfc.nasa.gov}.}) and the Simons Observatory\footnote{\tt{https://simonsobservatory.org.}} are going to deliver data at an unprecedented precision, allowing ever more stringent tests of $\Lambda$CDM. In order to interpret the results from these experiments, we require consistency measures between different data sets and methods that allow us to assess the goodness of fit of different models.

Several measures of concordance between constraints from different experiments have been proposed and used in the literature (see e.g. \cite{Hobson:2002, Kunz:2006, Marshall:2006, Verde:2013, Amendola:2013, Karpenka:2015, MacCrann:2015, Charnock:2017, Lin:2017, Adhikari:2018, Raveri:2018}). Examples of methods that allow us to compare different models include the Bayesian evidence (e.g \cite{Knuth:2015, Kass:1995}), the Akaike Information Criterion (AIC) \cite{Akaike:1974}, the Bayesian Information Criterion (BIC) \cite{Schwarz:1978} and the Deviance Information Criterion (DIC) \cite{Spiegelhalter:2002}. All these model comparison methods have been applied successfully to cosmology (e.g. \cite{Joudaki:2017, Heavens:2017, DES-Collaboration:2017}). In this work, we focus on the relative entropy or Kullback-Leibler (KL) divergence \cite{Kullback:1951}. The relative entropy has already been employed in Refs.~\cite{Seehars:2014, Seehars:2016, Grandis:2016, Grandis:2016aa, Zhao:2017, Nicola:2017aa} to assess consistency between several different data sets. In the first part of this paper, we revisit the relative entropy as a consistency measure, building on Refs.~\cite{Seehars:2014, Seehars:2016}. We investigate some of its key properties, such as asymmetry and path dependence, in more detail and illustrate them with a simple toy model. In a second part, we introduce a novel model rejection method, which is complementary to model comparison methods traditionally employed in cosmology. Our method is based on the KL divergence and the posterior predictive distribution (PPD), which has been used in Ref.~\cite{Feeney:2018} to quantify tensions in the value of the Hubble parameter. The proposed method and model rejection in general, has several convenient properties that make it a useful addition to existing approaches. These include that (i) our method, implementing model rejection, allows us to assess the goodness of fit of any model without considering an alternative, (ii) PPDs are generally close to Gaussian and (iii) consistency can be easily quantified with a $p$-value. We demonstrate the method on a series of toy models and apply it to Type Ia supernovae (SNe Ia) data from the JLA \cite{Betoule:2014} and Cosmic Microwave Background (CMB) constraints \cite{Planck-Collaboration:2016ai} in the framework of six different cosmological models. 

This paper is organized as follows. In Section \ref{sec:data_param_space}, we review the basic questions in inference analyses and the notion of data and parameter space. Section \ref{sec:rel_ent_consistency} revisits the relative entropy as a consistency measure, while Section \ref{sec:rel_ent_modelselection} introduces our novel model rejection method and describes an application to cosmological data. Finally, we conclude in Section \ref{sec:conclusions}. Robustness tests and implementation details are deferred to the Appendices.

\section{On questions and spaces in inference analyses}\label{sec:data_param_space}

Any inference problem involves two fundamental spaces: the space of all possible parameter values $\boldsymbol{\theta}_{i}$ of a given model $M$, i.e. model space $\mathcal{S}_{M}$, and the space of all possible outcomes $\textbf{y}_{i}$ of a given experiment $E$, i.e. data space $\mathcal{S}_{D}$. These two spaces are connected through an underlying, data-generating model, which we can use alongside Bayes' theorem to transform probability distributions between data and model space or vice versa. 

In connection to these distinct spaces, in any inference analysis, we can ask two separate questions: (i) Are the constraints from two different data sets consistent with each other within a given model (data set consistency)?, (ii) Given a set of different data-generating models, which ones are allowed/excluded by a given data set (model rejection)? Question (i) amounts to fixing a particular model and comparing model parameter constraints in $\mathcal{S}_{M}$ obtained with different data set combinations. Therefore, it is most easily approached in model space $\mathcal{S}_{M}$. In order to answer question (ii), we need to consider different model spaces, while the data remain the same. Therefore, the analysis cannot easily be performed in $\mathcal{S}_{M}$ as in case (i), but we rather have to test the respective models in data space $\mathcal{S}_{D}$.

In this work, we address questions (i) and (ii) using the relative entropy or KL divergence \cite{Kullback:1951}. The relative entropy $D_{\mathrm{KL}}$ between a prior $p_{1}(\boldsymbol{\theta})$ and a posterior distribution $p_{2}(\boldsymbol{\theta})$ is given by \cite{Kullback:1951}
\begin{equation}
D_{\mathrm{KL}}(p_{2}||p_{1}) = \int\mathrm{d}\boldsymbol{\theta} \: p_{2}(\boldsymbol{\theta}) \log \frac{p_{2}(\boldsymbol{\theta})}{p_{1}(\boldsymbol{\theta})}.
\label{eq:kldiv}
\end{equation}
It is a measure of the information gain for a Bayesian update from $p_{1}(\boldsymbol{\theta})$ to $p_{2}(\boldsymbol{\theta})$ and has units of bits, provided the logarithm in Eq.~\ref{eq:kldiv} is taken to base 2. 

\section{Data set consistency: fixed model and different data}\label{sec:rel_ent_consistency}

Comparing constraints obtained from different data sets in the framework of a fixed model is a useful test, as any tension is a sign for unaccounted-for systematics or problems with the assumed model. The relative entropy is a measure for the information gain for a Bayesian update from a probability distribution function (pdf) $p_{1}(\boldsymbol{\theta})$ to $p_{2}(\boldsymbol{\theta})$, and therefore does not allow assessing consistency between different data sets by itself. In order to extend the relative entropy into a consistency measure, Ref.~\cite{Seehars:2014} defined the Surprise statistic $S$ as the difference between the observed $D_{\mathrm{KL}}$ and the expected relative entropy $\langle D_{\mathrm{KL}} \rangle$, i.e. $S = D_{\mathrm{KL}} - \langle D_{\mathrm{KL}} \rangle$. By definition, the Surprise is expected to scatter around zero and significant deviations from this behavior allow us to detect inconsistencies: a large, positive value of $S$ suggests a tension between the considered data sets, as they are more different than expected. If on the other hand $S < 0$, the pdfs are more similar than expected. 

The Surprise statistic has been applied to a wide variety of data sets (e.g. \cite{Seehars:2014, Seehars:2016, Grandis:2016, Grandis:2016aa, Zhao:2017, Nicola:2017aa}). In the course of these works however, it has become clear that several properties of the relative entropy and the Surprise require further investigation and we aim to address these in the following.

\subsection{Sequential experiments}\label{sec:sequential_updating}

From Eq.~\ref{eq:kldiv}, we see that the relative entropy and thus the Surprise statistic are not symmetric upon exchange of prior and posterior\footnote{This implies that the relative entropy is mathematically not strictly a metric.}. When applying the KL divergence to testing the consistency of data sets $\textbf{y}_{1}$ and $\textbf{y}_{2}$\footnote{With this, we mean the consistency of the constraints on model parameters obtained from data sets $\textbf{y}_{1}$ and $\textbf{y}_{2}$, but in the following, we will use these two expressions interchangeably.}, obtained in two different experiments, we thus need to decide on a choice of prior and posterior. There exist two common cases in which this choice is rather clear: (i) when comparing an earlier experiment to a current one, it is natural to choose the distribution derived from the current experiment as posterior in $D_{\mathrm{KL}}$ and the distribution from the earlier experiment as the prior, and (ii) when comparing a more constraining to a less constraining experiment, it is equally natural to set the posterior to the distribution of the more constraining experiment and the prior to the less constraining one. In other situations however, the choice of prior and posterior is not obvious and the asymmetry of the relative entropy may appear as a problem. However, as we will see below, this asymmetry is a feature of the relative entropy, rather than a shortcoming.

\begin{figure*}
\begin{center}
\includegraphics[scale=0.45]{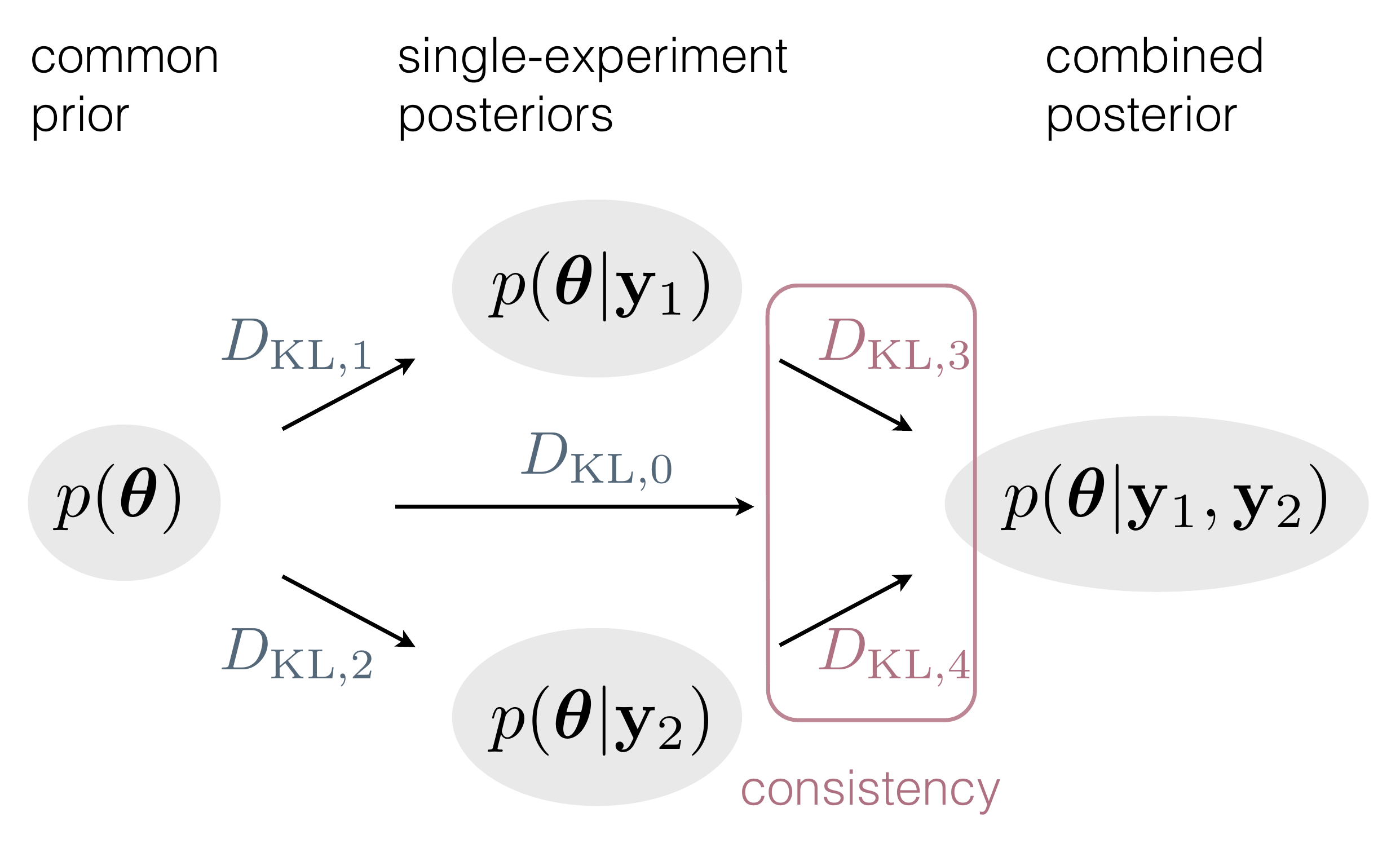}
\caption{Illustration of method to assess consistency of two experiments within a given model. The quantities $\textbf{y}_{1}$, $\textbf{y}_{2}$ denote the outcomes of the experiments and the KL divergences $D_{\mathrm{KL}, 1}, D_{\mathrm{KL}, 2}$ are the information gains for updates from a common prior $p(\boldsymbol{\theta})$ to the posteriors of the respective experiments. We quantify the consistency between the two data sets by computing the relative entropies for updates from the single-experiment posteriors $p_{1}(\boldsymbol{\theta}), p_{2}(\boldsymbol{\theta})$ to the combined posterior, i.e. $D_{\mathrm{KL}, 3}, D_{\mathrm{KL}, 4}$ and comparing them to expectations. If we find a significant difference between observed and expected relative entropy in either of these cases, we reject the null hypothesis that the data sets are consistent.}
\label{fig:diagram}
\end{center}
\end{figure*}

In order to investigate cases that do not fall in any of the two categories described above, we consider two uncorrelated experiments with comparable constraining power that cannot easily be arranged chronologically. The setup chosen is illustrated in Fig.~\ref{fig:diagram}. We assume that an identical model parameter prior $p(\boldsymbol{\theta})$ is used throughout the analysis, and denote the experimental outcomes as $\textbf{y}_{1}$ and $\textbf{y}_{2}$ respectively. The posterior derived from the first experiment is thus given by
\begin{equation}
p_{1}(\boldsymbol{\theta}) \equiv p(\boldsymbol{\theta} \vert \textbf{y}_{1}) = \frac{p(\boldsymbol{\theta})p(\textbf{y}_{1}\vert \boldsymbol{\theta})}{p(\textbf{y}_{1})},
\label{eq:posterior1}
\end{equation}
where $p(\textbf{y}_{1}\vert \boldsymbol{\theta})$ denotes the likelihood of the first experiment, $p(\boldsymbol{\theta})$ is the prior and the evidence is given by $p(\textbf{y}_{1})$. The posterior of the second experiment is analogously given by
\begin{equation}
p_{2}(\boldsymbol{\theta})\equiv p(\boldsymbol{\theta} \vert \textbf{y}_{2}) = \frac{p(\boldsymbol{\theta})p(\textbf{y}_{2}\vert \boldsymbol{\theta})}{p(\textbf{y}_{2})},
\label{eq:posterior2}
\end{equation}
where again $p(\textbf{y}_{2}\vert \boldsymbol{\theta})$ denotes the likelihood of the second experiment and the evidence is given by $p(\textbf{y}_{2})$. The information gains for a Bayesian update from prior $p(\boldsymbol{\theta})$ to posteriors $p_{1}(\boldsymbol{\theta})$, $p_{2}(\boldsymbol{\theta})$ respectively, are 
\begin{eqnarray}
D_{\mathrm{KL}, 1} \equiv D_{\mathrm{KL}}(p_{1}(\boldsymbol{\theta})||p(\boldsymbol{\theta})) = \int \mathrm{d}\boldsymbol{\theta} p_{1}(\boldsymbol{\theta}) \log{\frac{p_{1}(\boldsymbol{\theta})}{p(\boldsymbol{\theta})}}, \label{eq:dkl_pr1} \\
D_{\mathrm{KL}, 2} \equiv D_{\mathrm{KL}}(p_{2}(\boldsymbol{\theta})||p(\boldsymbol{\theta})) = \int \mathrm{d}\boldsymbol{\theta} p_{2}(\boldsymbol{\theta}) \log{\frac{p_{2}(\boldsymbol{\theta})}{p(\boldsymbol{\theta})}}. \label{eq:dkl_pr2}
\end{eqnarray}

We can further compute the information gains from $\textbf{y}_{2}$ and $\textbf{y}_{1}$ with respect to $p_{1}(\boldsymbol{\theta})$ and $p_{2}(\boldsymbol{\theta})$ taken as priors (see Fig.~\ref{fig:diagram}). As the two data sets are assumed to be uncorrelated, the combined posterior is given by the product of the prior and the likelihood of the second experiment, which gives (for both cases)
\begin{equation}
p_{12}(\boldsymbol{\theta})\equiv p(\boldsymbol{\theta} \vert \textbf{y}_{1}, \textbf{y}_{2}) = \frac{p(\boldsymbol{\theta}, \textbf{y}_{1}, \textbf{y}_{2})}{p(\textbf{y}_{1}, \textbf{y}_{2})} = \frac{p(\boldsymbol{\theta}\vert \textbf{y}_{1})p(\textbf{y}_{2} \vert \boldsymbol{\theta}, \textbf{y}_{1})}{p(\textbf{y}_{2}\vert \textbf{y}_{1})} = \frac{p(\boldsymbol{\theta}\vert \textbf{y}_{2})p(\textbf{y}_{1} \vert \boldsymbol{\theta}, \textbf{y}_{2})}{p(\textbf{y}_{1}\vert \textbf{y}_{2})}.
\label{eq:comb_post}
\end{equation}
Therefore, the associated relative entropies become
\begin{eqnarray}
D_{\mathrm{KL}, 3} \equiv D_{\mathrm{KL}}(p_{12}(\boldsymbol{\theta})||p_{1}(\boldsymbol{\theta})) = \int\mathrm{d}\boldsymbol{\theta} \: p_{12}(\boldsymbol{\theta}) \log \frac{p_{12}(\boldsymbol{\theta})}{p_{1}(\boldsymbol{\theta})}, \label{eq:dkl_1comb} \\
D_{\mathrm{KL}, 4} \equiv D_{\mathrm{KL}}(p_{12}(\boldsymbol{\theta})||p_{2}(\boldsymbol{\theta})) = \int\mathrm{d}\boldsymbol{\theta} \: p_{12}(\boldsymbol{\theta}) \log \frac{p_{12}(\boldsymbol{\theta})}{p_{2}(\boldsymbol{\theta})}. \label{eq:dkl_2comb}
\end{eqnarray}
In general, the relative entropies $D_{\mathrm{KL}, 3}$ and $D_{\mathrm{KL}, 4}$ are not equal, as the KL divergence depends both on the prior and the specific outcome of an experiment, $\textbf{y}_{i}$. This makes sense intuitively: if the prior is a delta function, i.e. we know the true value of the parameters, then no experiment can be informative \cite{Lindley:1956}. Likewise, depending on the outcome of an experiment, we can gain more or less information; if, for example, an experiment results in a very unlikely outcome, then the uncertainty on the value of the parameter might be larger after the experiment has been performed \cite{Lindley:1956, Caticha:2008}. This implies that the relative entropy depends on the path taken to reach a given result and is thus not additive for sequential updates\footnote{This means that in general we have $D_{\mathrm{KL}, 0} \neq D_{\mathrm{KL}, 1} + D_{\mathrm{KL}, 3}$ or $D_{\mathrm{KL}, 0} \neq D_{\mathrm{KL}, 2} + D_{\mathrm{KL}, 4}$.}. We can eliminate the dependence of the KL divergence on $\textbf{y}_{i}$ by averaging over all possible experimental outcomes. This was done in Ref.~\cite{Seehars:2014} to derive the expected relative entropy, which for $D_{\mathrm{KL}, 1}$ as an example, is given by:
\begin{equation}
\langle D_{\mathrm{KL}}(p_{1}(\boldsymbol{\theta})||p(\boldsymbol{\theta})) \rangle = \int \mathrm{d}\textbf{y}_{1} p(\textbf{y}_{1}) D_{\mathrm{KL}}(p_{1}(\boldsymbol{\theta})||p(\boldsymbol{\theta})), 
\label{eq:exp_rel_ent}
\end{equation}
where $p(\textbf{y}_{1})$ is the prior predictive distribution for $\textbf{y}_{1}$\footnote{The prior predictive distribution, evaluated for the observed data, is equivalent to the evidence.}, defined as
\begin{equation}
p(\textbf{y}_{1}) = \int \mathrm{d}\boldsymbol{\theta} p(\textbf{y}_{1}\vert \boldsymbol{\theta}) p(\boldsymbol{\theta}).
\end{equation} 
As shown in Appendix \ref{ap:path_dependence}, the expected relative entropy turns out to be additive for sequential updates. The observed KL divergence and thus the Surprise on the other hand, do not add and it is important to explicitly take the path into account when assessing the consistency between two arbitrary experiments with outcomes $\textbf{y}_{1}$ and $\textbf{y}_{2}$. Therefore, we propose to compute both $D_{\mathrm{KL}, 3}$ and $D_{\mathrm{KL}, 4}$, as well as the associated Surprise values, $S_{3}$, $S_{4}$. If both $S_{3}$ and $S_{4}$ are compatible with $0$, the two data sets are consistent. In the case in which one or both Surprise values are significantly larger than expected, we propose to reject the null hypothesis of consistency between $\textbf{y}_{1}$ and $\textbf{y}_{2}$, as the path dependence of the KL divergence can lead to inconsistencies only being detectable in one of the two updates. In the following, we illustrate this with a toy model. For an alternative discussion of the case in which $S_{3}$ and $S_{4}$ give inconsistent results, the reader in referred to Appendix ~\ref{ap:dkl_asymmetry}\footnote{We note that analogous considerations can also be made if $\textbf{y}_{1}$ and $\textbf{y}_{2}$ are correlated, with the only difference that the combined posterior in Eq.~\ref{eq:comb_post} is given by the joint analysis of the two data sets.}.

\begin{figure}
\begin{center}
\includegraphics[width=0.95\textwidth]{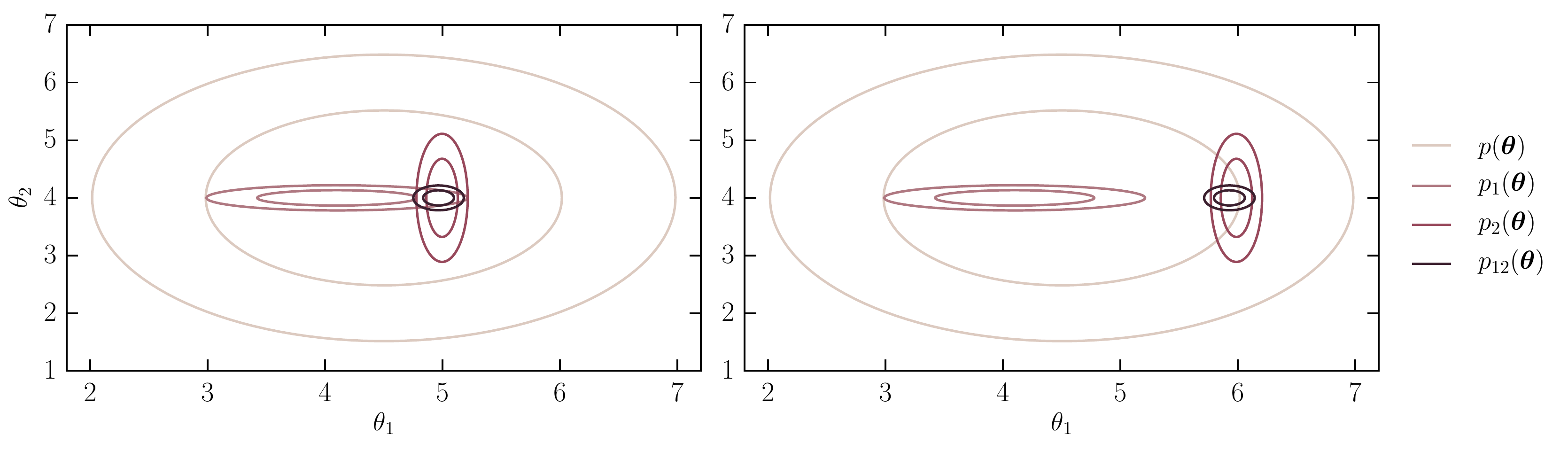}
\caption{Illustration of the probability distributions in parameter space considered in the toy models of Sec.~\ref{subsec:toy_model_asymmetry}; on the left, toy model I; on the right, toy model II. In each case the inner (outer) contour shows the $68 \%$ confidence limits (c.l.) ($95 \%$ c.l.).}
\label{fig:toy-model}
\end{center}
\end{figure}

\subsection{Toy model}\label{subsec:toy_model_asymmetry}

In order to illustrate the concepts discussed above, we consider one of the toy models introduced in Ref.~\cite{Lin:2017}. This toy model consists of two uncorrelated, contemporaneous experiments with likelihoods $p(\textbf{y}_{1}\vert \boldsymbol{\theta})$ and $p(\textbf{y}_{2}\vert \boldsymbol{\theta})$. We complement this toy model with a common prior $p(\boldsymbol{\theta})$ and the configuration (toy model I) is illustrated in the left panel of Fig.~\ref{fig:toy-model}. Both likelihoods $p(\textbf{y}_{1}\vert \boldsymbol{\theta})$ and $p(\textbf{y}_{2}\vert \boldsymbol{\theta})$ are assumed to be Gaussian with means $\boldsymbol{\mu}_{i}$ and covariances $C_{i}$ 
\begin{eqnarray}
p(\textbf{y}_{1}\vert \boldsymbol{\theta}): \; \boldsymbol{\mu}_{1} = \big(4, 4\big), \; C_{1} = \begin{pmatrix} \sfrac{1}{4} & 0 \\ 0 & \sfrac{1}{128} \end{pmatrix}, \\ 
p(\textbf{y}_{2}\vert \boldsymbol{\theta}): \; \boldsymbol{\mu}_{2} = \big(5, 4\big), \; C_{2} = \begin{pmatrix} \sfrac{1}{128} & 0 \\ 0 & \sfrac{1}{4} \end{pmatrix}.
\end{eqnarray}
Furthermore, we assume a wide, Gaussian prior $p(\boldsymbol{\theta})$, given by
\begin{equation}
p(\boldsymbol{\theta}): \; \boldsymbol{\mu}_{p} = \big(4.5, 4\big), \; C_{p} = \begin{pmatrix} 1 & 0 \\ 0 & 1 \end{pmatrix}.
\end{equation}
We quantify the consistency between $\textbf{y}_{1}$ and $\textbf{y}_{2}$ by computing the relative entropy and the Surprise between the separately derived posteriors and their combination (also see Fig.~\ref{fig:diagram}). As all distributions involved are Gaussian, we compute the KL divergences analytically \cite{Seehars:2014} using the publicly available $\texttt{Surprise}$ package\footnote{The package can be found at $\texttt{https://github.com/seeh/surprise}$.}, described in Ref.~\cite{Seehars:2016}. The obtained results are given in Tab.~\ref{tab:relent}. 

\begin{table*}
\caption{Values of the relative entropy $D_{\mathrm{KL}}$, expected relative entropy $\langle D_{\mathrm{KL}} \rangle$, standard deviation of the relative entropy $\sigma(D_{\mathrm{KL}})$ and Surprise $S$ for the parameter updates considered in the toy model. All values are given in units of bits. The $p$-values denote the probability of observing a Surprise value greater or equal (less or equal) than $S$ if $S$ is greater (smaller) than zero, when assuming consistency between the two data sets.} \label{tab:relent}
\begin{center}
\begin{tabular}{c>{\raggedright}m{0.8cm}>{\raggedright}m{0.5cm}c>{\raggedright}m{0.8cm}ccccc}
\hline\hline 
Toy model & \multicolumn{4}{l}{Data combination} & $D_{\mathrm{KL}}$ & $\langle D_{\mathrm{KL}} \rangle$ & $S$ & $\sigma(D_{\mathrm{KL}})$ & $p$-value \\ \hline      
\multirow{5}{*}{I} & $D_{\mathrm{KL}, 1}$: & $p(\boldsymbol{\theta})$ & $\rightarrow$ & $p_{1}(\boldsymbol{\theta})$ & 3.5 & 4.7 & -1.2 & 1.3 & 0.08 \\
& $D_{\mathrm{KL}, 2}$: & $p(\boldsymbol{\theta})$ & $\rightarrow$ & $p_{2}(\boldsymbol{\theta})$ & 3.6 & 4.7 & -1.1 & 1.3 & 0.13 \\
& $D_{\mathrm{KL}, 3}$: & $p_{1}(\boldsymbol{\theta})$ & $\rightarrow$ & $p_{12}(\boldsymbol{\theta})$ & 4.4 & 2.4 & 2.0 & 1.0 & 0.05 \\
& $D_{\mathrm{KL}, 4}$: & $p_{2}(\boldsymbol{\theta})$ & $\rightarrow$ & $p_{12}(\boldsymbol{\theta})$ & 1.8 & 2.4 & -0.6 & 1.0 & 0.25 \\
& $D_{\mathrm{KL}, 0}$: & $p(\boldsymbol{\theta})$ & $\rightarrow$ & $p_{12}(\boldsymbol{\theta})$ & 5.8 & 7.1 & -1.3 & 1.4 & 0.1 \\\hdashline
\multirow{5}{*}{II} & $D_{\mathrm{KL}, 1}$: & $p(\boldsymbol{\theta})$ & $\rightarrow$ & $p_{1}(\boldsymbol{\theta})$ & 3.5 & 4.7 & -1.2 & 1.3 & 0.08 \\
& $D_{\mathrm{KL}, 2}$: & $p(\boldsymbol{\theta})$ & $\rightarrow$ & $p_{2}(\boldsymbol{\theta})$ & 5.0 & 4.7 & 0.3 & 1.3 & 0.29 \\
& $D_{\mathrm{KL}, 3}$: & $p_{1}(\boldsymbol{\theta})$ & $\rightarrow$ & $p_{12}(\boldsymbol{\theta})$ & 13.8 & 2.4 & 11.4 & 1.0 & $3 \times 10^{-5}$ \\
& $D_{\mathrm{KL}, 4}$: & $p_{2}(\boldsymbol{\theta})$ & $\rightarrow$ & $p_{12}(\boldsymbol{\theta})$ & 2.0 & 2.4 & -0.4 & 1.0 & 0.50 \\
& $D_{\mathrm{KL}, 0}$: & $p(\boldsymbol{\theta})$ & $\rightarrow$ & $p_{12}(\boldsymbol{\theta})$ & 7.1 & 7.1 & 0.0 & 1.4 & 0.6 \\
 \hline \hline
\end{tabular}
\end{center}
\end{table*}

From Tab.~\ref{tab:relent}, we first of all see that the observed relative entropies are not additive, while the expected information gains are, which is expected from the discussions in Sec.~\ref{sec:sequential_updating} and Appendix \ref{ap:path_dependence}. Furthermore, we find that the Surprises $S_{3}, S_{4}$ obtained for $D_{\mathrm{KL}, 3}$ and $D_{\mathrm{KL}, 4}$ differ in both magnitude and sign. The first case shows a positive, marginally significant Surprise, while the second case shows a negative Surprise, suggesting that $p_{1}(\boldsymbol{\theta})$ and $p_{12}(\boldsymbol{\theta})$ are slightly more different than expected, while $p_{2}(\boldsymbol{\theta})$ and $p_{12}(\boldsymbol{\theta})$ are more similar than expected. This is also evident from the left panel of Fig.~\ref{fig:toy-model}: the means of the combined posterior and $p_{1}(\boldsymbol{\theta})$ differ by more than one standard deviation of the latter distribution, while the means of $p_{2}(\boldsymbol{\theta})$ and $p_{12}(\boldsymbol{\theta})$ are very similar. This asymmetry thus explains the difference in observed relative entropies. 

However, as neither $D_{\mathrm{KL}, 3}$ or $D_{\mathrm{KL}, 4}$ suggest a significant tension between the two data sets, we conclude that $\textbf{y}_{1}$ and $\textbf{y}_{2}$ are consistent\footnote{We use the convention of $p$-value $< 2.7 \times 10^{-3}$ as our definition of significant discrepancy, which corresponds to a Gaussian equivalent of $3\sigma$.}. We can construct a more discrepant case (toy model II) by moving the mean of $p(\textbf{y}_{2} \vert \boldsymbol{\theta})$ to $\boldsymbol{\mu}_{2} = \big(6, 4 \big)$, as shown in the right panel of Fig.~\ref{fig:toy-model}. In this case, we find a significant Surprise for $D_{\mathrm{KL}, 3}$ ($S_{3} = 11.4$, $\sigma(D_{\mathrm{KL}, 3}) = 1.0$ bits), while $p_{2}(\boldsymbol{\theta})$ and $p_{12}(\boldsymbol{\theta})$ are still consistent (see Tab.~\ref{tab:relent}). This is equivalent to the results obtained for toy model I except that the discrepancy in $D_{\mathrm{KL}, 3}$ is more pronounced. We thus see that inconsistencies in parameters in which the constraining power of the two distributions is different, tend to only appear in the update from the weaker constraints to the combined distribution. The reason is that the combined mean lies close to the mean of the distribution with the stronger constraining power and thus the latter two will be consistent even though the original distributions are not\footnote{This will cease to be true for large discrepancies, as the mean of the more constraining distribution will then shift as well.}. In such situations, the asymmetry of the relative entropy therefore additionally allows us to determine the distribution that is most likely to drive the discrepancy (in this case the more constraining one).

This toy model shows that a significant discrepancy in at least one of the updates is a sign for a tension between the distributions being combined. It is thus advisable to consider both updates (e.g. $D_{\mathrm{KL}, 3}$, $D_{\mathrm{KL}, 4}$ in this case) for concurrent experiments with equal constraining power, and reject the null hypothesis of consistency if any of the two updates shows a significant tension.  

\section{Model rejection: fixed data and different models}\label{sec:rel_ent_modelselection}

In addition to assessing the consistency between different data sets, model rejection provides a complementary way to further improve our understanding of the standard cosmological model. In a model rejection analysis, we determine the goodness of fit of a given model to the data. This analysis neither requires the comparison between different data nor an alternative model. As discussed in Sec.~\ref{sec:data_param_space}, data space, $\mathcal{S}_{D}$, is fixed in model rejection, while we compare different model spaces, $\mathcal{S}_{M}$. Therefore, it appears simpler to compare different models in data space $\mathcal{S}_{D}$, although other approaches are possible and exist in the literature. In this work, we choose to work in data space and thus the first step towards model rejection with the relative entropy is to identify appropriate priors and posteriors in $\mathcal{S}_{D}$.

A central quantity in Bayesian model testing is the posterior predictive distribution (see e.g. \cite{Gelman:1996}). The PPD quantifies the probability of future data $\textbf{y}^{\prime}$ conditioned on the current observation $Y = \textbf{y}$, given an underlying model $M$ that links data and model space (for an application of the PPD in cosmology, see \cite{Feeney:2018}). Essentially, the PPD is the average of the likelihood of the new data under the posterior of the model parameters, i.e.
\begin{equation}
p(\textbf{y}^{\prime} \vert \textbf{y}, C_{\textbf{y}}, C_{\textbf{y}^{\prime}}, M) = \int \mathrm{d}\boldsymbol{\theta} p(\textbf{y}^{\prime} \vert \boldsymbol{\theta}, C_{\textbf{y}^{\prime}}) p(\boldsymbol{\theta} \vert \textbf{y}, C_{\textbf{y}}),
\label{eq:eq_ppd}
\end{equation}
where $\boldsymbol{\theta}$ denotes the vector of model parameters of $M$ and $C_{\textbf{y}}, C_{\textbf{y}^{\prime}}$ are the covariance matrices of current and future data respectively. Furthermore, $p(\boldsymbol{\theta} \vert \textbf{y}, C_{\textbf{y}})$ denotes the posterior distribution for the parameters $\boldsymbol{\theta}$ and $p(\textbf{y}^{\prime} \vert \boldsymbol{\theta}, C_{\textbf{y}^{\prime}})$ is the likelihood of the new data, which does not necessarily equal the likelihood of the current data. In Eq.~\ref{eq:eq_ppd} we explicitly condition on a model $M$. We can however equally define a PPD conditioned only on current data. This latter quantity is given by
\begin{equation}
p(\textbf{y}^{\prime} \vert \textbf{y}, C_{\textbf{y}}, C_{\textbf{y}^{\prime}}) = \int \mathrm{d}\hat{\textbf{y}}_{\mathrm{t}} p(\textbf{y}^{\prime} \vert \hat{\textbf{y}}_{\mathrm{t}}, C_{\textbf{y}^{\prime}}) p(\hat{\textbf{y}}_{\mathrm{t}} \vert \textbf{y}, C_{\textbf{y}}),
\label{eq:eq_ppd_data}
\end{equation}
where $\hat{\textbf{y}}_{\mathrm{t}}$ denotes the expectation for the underlying data, i.e. the data points that would be measured in an infinite precision experiment. The quantity $p(\hat{\textbf{y}}_{\mathrm{t}} \vert \textbf{y}, C_{\textbf{y}})$ is the sampling distribution of $\hat{\textbf{y}}_{\mathrm{t}}$ given current data and $p(\textbf{y}^{\prime} \vert \hat{\textbf{y}}_{\mathrm{t}}, C_{\textbf{y}^{\prime}})$ is the likelihood of future data conditioned on $\hat{\textbf{y}}_{\mathrm{t}}$. The sampling distribution of $\hat{\textbf{y}}_{\mathrm{t}}$ can be thought of as the expectation for $\hat{\textbf{y}}_{\mathrm{t}}$ given observation $\textbf{y}$ and covariance $C_{\textbf{y}}$: If we believe both our measurement and its covariance, we expect $\hat{\textbf{y}}_{\mathrm{t}}$ to scatter around $\textbf{y}$, consistent with $C_{\textbf{y}}$. For a similar approach, also see Ref.~\cite{Amara:2014}.

Eq.~\ref{eq:eq_ppd_data} expresses the distribution of future data as predicted using only current data. We call this the PPD from the data, while we denote Eq.~\ref{eq:eq_ppd} as PPD from data and model. 

\begin{figure}
\begin{center}
\includegraphics[scale=0.45]{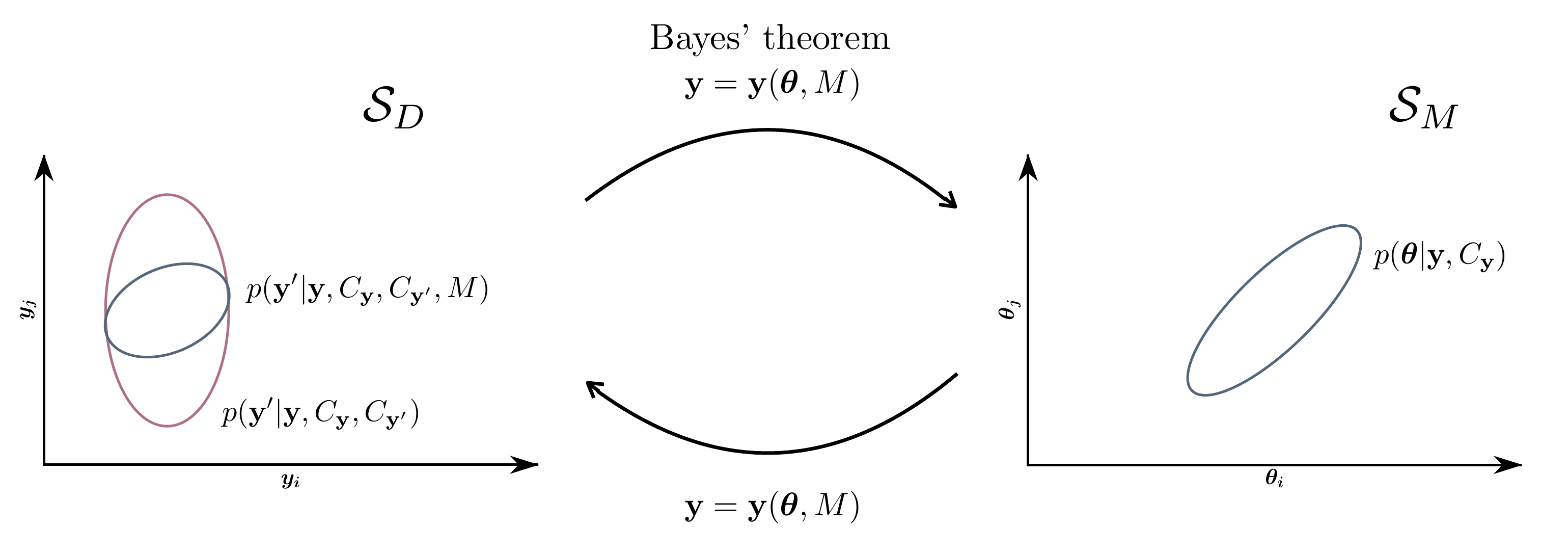}
\caption{Illustration of model rejection framework proposed in this work. The left panel shows data space $\mathcal{S}_{D}$, while model space $\mathcal{S}_{M}$ is illustrated in the right panel. The mapping from $\mathcal{S}_{D}$ to $\mathcal{S}_{M}$ is provided by Bayes' theorem and the data-generating model, $\textbf{y} = \textbf{y}(\boldsymbol{\theta}, M)$. To quantify how well a model fits the data, we compute the relative entropy between the PPD derived from data and model $p(\textbf{y}^{\prime} \vert \textbf{y}, C_{\textbf{y}}, C_{\textbf{y}^{\prime}}, M) \in \mathcal{S}_{D}$ and the PPD derived only from the data $p(\textbf{y}^{\prime} \vert \textbf{y}, C_{\textbf{y}}, C_{\textbf{y}^{\prime}}) \in \mathcal{S}_{D}$. A relative entropy significantly larger than expected signifies that the model is ruled out, while consistency between expected and observed KL divergence means that the model is allowed by the data.}
\label{fig:ppds}
\end{center}
\end{figure}

If model $M$ captures the relevant features in the data, we expect the two distributions in Equations \ref{eq:eq_ppd} and \ref{eq:eq_ppd_data} to be similar, as the observed data $\textbf{y}$ represent a typical draw from the underlying model. If on the other hand, $M$ is not a good fit to the data, we expect to see significant differences, especially in the means of the distributions. 

In order to assess how well model $M$ describes the data, we therefore propose to quantify the difference between $p(\textbf{y}^{\prime} \vert \textbf{y}, C_{\textbf{y}}, C_{\textbf{y}^{\prime}})$ and $p(\textbf{y}^{\prime} \vert \textbf{y}, C_{\textbf{y}}, C_{\textbf{y}^{\prime}}, M)$ through their relative entropy $D_{\mathrm{KL}}$, by considering
\begin{equation}
D_{\mathrm{KL}}(p(\textbf{y}^{\prime} \vert \textbf{y}, C_{\textbf{y}}, C_{\textbf{y}^{\prime}}, M) || p(\textbf{y}^{\prime} \vert \textbf{y}, C_{\textbf{y}}, C_{\textbf{y}^{\prime}})) = \int \mathrm{d}\textbf{y}^{\prime} p(\textbf{y}^{\prime} \vert \textbf{y}, C_{\textbf{y}}, C_{\textbf{y}^{\prime}}, M)  \log{\frac{p(\textbf{y}^{\prime} \vert \textbf{y}, C_{\textbf{y}}, C_{\textbf{y}^{\prime}}, M)}{p(\textbf{y}^{\prime} \vert \textbf{y}, C_{\textbf{y}}, C_{\textbf{y}^{\prime}})}},
\label{eq:eq_dkl}
\end{equation}
where we have set the prior to the PPD from the data and the posterior to the PPD from the data and model (for an illustration, see Fig.~\ref{fig:ppds}). Note that this is similar to the model breaking figure of merit introduced in Ref.~\cite{Amara:2014}. With this choice, Eq.~\ref{eq:eq_dkl} can be interpreted as the information gain coming from the assumption of a given model $M$. Analogously to the discussion in Sec.~\ref{sec:rel_ent_consistency}, the value of $D_{\mathrm{KL}}$ by itself does not allow us to quantify the consistency between data and model $M$. In analogy to the Surprise statistic \cite{Seehars:2014, Seehars:2016}, we need to compare the observed relative entropy to the relative entropy expected under the null hypothesis, which we denote by $\langle D_{\mathrm{KL}} \rangle$. In the present case, the null hypothesis states that the data are drawn from model $M$ and we can thus define the expected relative entropy as
\begin{multline}
\langle D_{\mathrm{KL}}(p(\textbf{y}^{\prime}(M) \vert \textbf{y}(M), C_{\textbf{y}}, C_{\textbf{y}^{\prime}}, M) || p(\textbf{y}^{\prime}(M) \vert \textbf{y}(M), C_{\textbf{y}}, C_{\textbf{y}^{\prime}})) \rangle = \\ \iint \mathrm{d}\textbf{y}(M) \; \mathrm{d}\textbf{y}^{\prime}(M) \; p(\textbf{y}(M) \vert C_{\textbf{y}}, M) p(\textbf{y}^{\prime}(M) \vert \textbf{y}(M), C_{\textbf{y}}, C_{\textbf{y}^{\prime}}, M)  \log{\frac{p(\textbf{y}^{\prime}(M) \vert \textbf{y}(M), C_{\textbf{y}}, C_{\textbf{y}^{\prime}}, M)}{p(\textbf{y}^{\prime}(M) \vert \textbf{y}(M), C_{\textbf{y}}, C_{\textbf{y}^{\prime}})}},
\label{eq:eq_dkl_exp}
\end{multline}
where $\textbf{y}(M)$ implies that $\textbf{y}$ is drawn from model $M$ and $p(\textbf{y}(M) \vert C_{\textbf{y}}, M)$ denotes the pdf of $\textbf{y}(M)$ given $M$. We outline the procedure to compute this quantity in Sec.~\ref{subsec:implementation}. To simplify the notation, we further define:
\begin{align}
D_{\mathrm{KL}}(\textbf{y}, M || \textbf{y}) &\equiv D_{\mathrm{KL}}(p(\textbf{y}^{\prime} \vert \textbf{y}, C_{\textbf{y}}, C_{\textbf{y}^{\prime}}, M) || p(\textbf{y}^{\prime} \vert \textbf{y}, C_{\textbf{y}}, C_{\textbf{y}^{\prime}})), \\
\langle D_{\mathrm{KL}}(\textbf{y}(M), M || \textbf{y}(M)) \rangle &\equiv \langle D_{\mathrm{KL}}(p(\textbf{y}^{\prime}(M) \vert \textbf{y}(M), C_{\textbf{y}}, C_{\textbf{y}^{\prime}}, M) || p(\textbf{y}^{\prime}(M) \vert \textbf{y}(M), C_{\textbf{y}}, C_{\textbf{y}^{\prime}})) \rangle. 
\end{align}

These two quantities, combined with the standard deviation of the expected relative entropy, $\sigma(\langle D_{\mathrm{KL}}(\textbf{y}(M), M || \textbf{y}(M)) \rangle)$, allow us to assess how well $M$ describes the data: a relative entropy $D_{\mathrm{KL}}(\textbf{y}, M || \textbf{y})$ significantly larger than $\langle D_{\mathrm{KL}}(\textbf{y}(M), M || \textbf{y}(M)) \rangle$ suggests that $M$ is not a good fit to the data, as the two PPDs are significantly different. Consistency between $D_{\mathrm{KL}}(\textbf{y}, M || \textbf{y})$ and $\langle D_{\mathrm{KL}}(\textbf{y}(M), M || \textbf{y}(M)) \rangle$ on the other hand, means that we cannot rule out the null hypothesis and model $M$ thus provides a good fit to the data. 

In this work, we compute the quantities $D_{\mathrm{KL}}(\textbf{y}, M || \textbf{y})$, $\langle D_{\mathrm{KL}}(\textbf{y}(M), M || \textbf{y}(M)) \rangle$ and \\ $\sigma(\langle D_{\mathrm{KL}}(\textbf{y}(M), M || \textbf{y}(M)) \rangle)$ assuming all distributions considered to be well-approximated by Gaussians, which allows us to compute all relative entropies analytically using the expressions given in Ref.~\cite{Seehars:2014}. This approximation is trivially justified for the toy model considered in Sec.~\ref{subsec:toy_model_model_selection}. In Sec.~\ref{subsec:cosmology_applic}, we apply the model rejection method to cosmological data from SNe Ia and CMB and we test the Gaussianity of the relevant PPDs in Appendix \ref{ap:gaussianity_tests}. We find them to be well-approximated by normal distributions, which is due to the fact that PPDs are generally close to Gaussian if the data is normally distributed, regardless of the posteriors.

In this work, we focus on developing a method that allows us to reject, rather than compare models. Model comparison can for example be performed using the Bayes' ratio, i.e. the evidence ratio of two different models (see e.g. \cite{Knuth:2015}). The Bayes' ratio includes an Occam penalty, which means that simpler models are generally preferred over complex ones. This makes sense intuitively, as increasing the number of degrees of freedom of a model $M$ will enable us to fit an increasing number of features in the data. Thus, growing model complexity generally leads to predictions closer to the specific data realization obtained, which means that the model will start fitting the noise in addition to the signal. The method proposed in this work does not penalize complex over simpler models, but we believe it possible to extend it to also include an Occam penalty. We leave an investigation thereof to future work.


\subsection{Implementation}\label{subsec:implementation}

We estimate all quantities discussed in the previous section using simple Monte Carlo simulations. For a given, data-generating model $M$ and observed data $\textbf{y}$ with likelihood $p(\textbf{y} \vert \boldsymbol{\theta}, C_{\textbf{y}})$, we first derive the posterior distribution for the model parameters $p(\boldsymbol{\theta} \vert \textbf{y}, C_{\textbf{y}})$ and the sampling distribution of $\hat{\textbf{y}}_{\mathrm{t}}$, $p(\hat{\textbf{y}}_{\mathrm{t}} \vert \textbf{y}, C_{\textbf{y}})$. The method for computing the observed relative entropy (Eq.~\ref{eq:eq_dkl}) using the integral expressions in Equations \ref{eq:eq_ppd} and \ref{eq:eq_ppd_data}, can then be summarized as:
\begin{enumerate}[label=(\roman*)]
\item Eq.~\ref{eq:eq_ppd}: For $i=1,\ldots, N$, first sample $\boldsymbol{\theta}_{i}$ from $p(\boldsymbol{\theta} \vert \textbf{y}, C_{\textbf{y}})$, then sample a realization $\textbf{y}^{\prime}_{i}$ of length $\bar{N}$ from $p(\textbf{y}^{\prime} \vert \boldsymbol{\theta}, C_{\textbf{y}^{\prime}})$. Marginalizing over $\boldsymbol{\theta}$, i.e. combining all the samples $\textbf{y}^{\prime}_{i}, i=1,\ldots, N$ gives us a sample from $p(\textbf{y}^{\prime} \vert \textbf{y}, C_{\textbf{y}}, C_{\textbf{y}^{\prime}}, M)$. 
\item Eq.~\ref{eq:eq_ppd_data}: For $i=1,\ldots, N$, first sample $\hat{\textbf{y}}_{\mathrm{t}, i}$ from $p(\hat{\textbf{y}}_{\mathrm{t}} \vert \textbf{y}, C_{\textbf{y}})$, then sample a realization $\textbf{y}^{\prime}_{i}$ of length $\bar{N}$ from $p(\textbf{y}^{\prime} \vert \hat{\textbf{y}}_{\mathrm{t}}, C_{\textbf{y}^{\prime}})$. Marginalizing over $\hat{\textbf{y}}_{\mathrm{t}}$, i.e. combining all the samples $\textbf{y}^{\prime}_{i}, i=1,\ldots, N$ gives us a sample from $p(\textbf{y}^{\prime} \vert \textbf{y}, C_{\textbf{y}}, C_{\textbf{y}^{\prime}})$. 
\end{enumerate}
In a last step we use these two combined samples to compute the relative entropy defined in Eq.~\ref{eq:eq_dkl}. For an alternative method, which can be applied if we can sample from the PPDs directly, see Appendix \ref{ap:alternat_rel_ent_alg}.

Recall that we have defined the expected relative entropy in Eq.~\ref{eq:eq_dkl_exp} as the information gain from the model $M$ conditional on the data being drawn from that same model. Instead of attempting to analytically predict this quantity, we again resort to a Monte Carlo approach. For each model $M$ considered in our analysis, we first derive the best-fit model parameters from the data. Using these best-fit parameters we generate a mock true data set $\textbf{y}_{\mathrm{t}}$. We then sample a realization $\textbf{y}$ from the true data distribution $p(\textbf{y} \vert \textbf{y}_{\mathrm{t}}, C_{\textbf{y}_{\mathrm{t}}})$, compute the best-fit values of the parameters of model $M$ and the corresponding posterior $p(\boldsymbol{\theta} \vert \textbf{y}, C_{\textbf{y}})$. For this simulated data set we now compute the relative entropy $D_{\mathrm{KL}}(\textbf{y}, M || \textbf{y})$ using a procedure analogous to the one employed for the real data: we sample realizations from Equations \ref{eq:eq_ppd} and \ref{eq:eq_ppd_data}, except that we have now made sure that the simulated data is explicitly drawn from an underlying model. Repeating this experiment $N$ times for different realizations $\textbf{y}$ of $\textbf{y}_{\mathrm{t}}$ finally gives us a sample of $D_{\mathrm{KL}}(\textbf{y}(M), M || \textbf{y}(M))$ and we estimate $\langle D_{\mathrm{KL}}(\textbf{y}(M), M || \textbf{y}(M)) \rangle$ and $\sigma(\langle D_{\mathrm{KL}}(\textbf{y}(M), M || \textbf{y}(M)) \rangle)$ as the mean and the standard deviation of this distribution respectively. 

The three quantities $D_{\mathrm{KL}}(\textbf{y}, M || \textbf{y})$, $\langle D_{\mathrm{KL}}(\textbf{y}(M), M || \textbf{y}(M)) \rangle$ and $\sigma(\langle D_{\mathrm{KL}}(\textbf{y}(M), M || \textbf{y}(M)) \rangle)$ allow us to assess the consistency of each model $M$ with the observed data. In practice, we use the full distribution of $D_{\mathrm{KL}}(\textbf{y}(M), M || \textbf{y}(M))$ to compute the one-sided $p$-value, i.e. the probability of observing a value greater or equal than $D_{\mathrm{KL}}(\textbf{y}, M || \textbf{y})$ under the null hypothesis that the data is drawn from model $M$\footnote{For comparison, when applying this method to cosmological data, we also compute $p$-values through the realized discrepancy algorithm of Ref.~\cite{Gelman:1996}. We find that both methods yield comparable results; for implementation details, the reader is referred to Appendix \ref{ap:realized_disc}.}. We note that there exist many measures different from the $p$-value to assess consistency; for an alternative example, see e.g. Ref.~\cite{Feeney:2018}. In this work, we have chosen the $p$-value, as it is obtained from an integral over the distribution function, which can easily be computed from the full distribution of $D_{\mathrm{KL}}(\textbf{y}(M), M || \textbf{y}(M))$ derived here.

\subsection{Toy model}\label{subsec:toy_model_model_selection}

We illustrate the methods described in the previous section by applying them to a toy model in which we consider fitting data with a polynomial of varying order. For illustration, we consider polynomials of order $n = 0 \ldots 7$, where $n=0$ denotes a linear polynomial with $y$-intercept set to zero. For each polynomial of degree $n$ we choose fiducial coefficients and generate simulated data drawn from this model with covariance matrix $C_{\textbf{y}}$. We then fit this data with polynomials of varying degree $\hat{n}$ and assess the consistency of model and data by computing the KL divergences discussed above. For each fiducial polynomial model of degree $n$, the procedure can thus be summarized as follows:
\begin{enumerate}
\item Draw a data realization $\textbf{y}$ from a fiducial polynomial $p_{n} \in \mathscr{P}_{n}$, where $\mathscr{P}_{n}$ denotes the vector space of all polynomials of degree $n$. We denote this polynomial as model $M_{n}$.
\item Fit $\textbf{y}$ with a polynomial $p_{\hat{n}} \in \mathscr{P}_{\hat{n}}$ with $\hat{n} \in [0, n_{\mathrm{max}}]$, where $\hat{n}$ is not necessarily equal to $n$ and determine the corresponding posterior $p(\boldsymbol{\theta} \vert \textbf{y}, C_{\textbf{y}})$. We denote this polynomial as model $M_{\hat{n}}$.
\item Compute the observed relative entropy $D_{\mathrm{KL}}(\textbf{y}, M_{\hat{n}} || \textbf{y})$, the expected relative entropy $\langle D_{\mathrm{KL}}(\textbf{y}(M_{\hat{n}}), M_{\hat{n}} || \textbf{y}(M_{\hat{n}})) \rangle$ and the standard deviation of the relative entropy \\ $\sigma(\langle D_{\mathrm{KL}}(\textbf{y}(M_{\hat{n}}), M_{\hat{n}} || \textbf{y}(M_{\hat{n}})) \rangle)$. Note that the expected relative entropy has to be computed for each model $M_{\hat{n}}$ separately.
\item Compute the $p$-value of the observed $D_{\mathrm{KL}}(\textbf{y}, M_{\hat{n}} || \textbf{y})$ for the pdf of the expected relative entropy.
\end{enumerate}
In our implementation, we assume the data to be drawn from a Gaussian distribution with covariance matrix $C_{\textbf{y}} = \sigma^{2}\mathbb{1}$ and we generate $N = 100$ realizations of length $\bar{N} = 1000$ each to compute the KL divergences $D_{\mathrm{KL}}(\textbf{y}, M_{\hat{n}} || \textbf{y})$ and $\langle D_{\mathrm{KL}}(\textbf{y}(M_{\hat{n}}), M_{\hat{n}} || \textbf{y}(M_{\hat{n}})) \rangle$. For a detailed description of how we determine the fiducial polynomial coefficients and the specific implementation of the toy model, the reader is referred to Appendix \ref{ap:toy_model_implementation}.

\begin{figure}
\begin{center}
\subfigure{\includegraphics[width=0.49\textwidth]{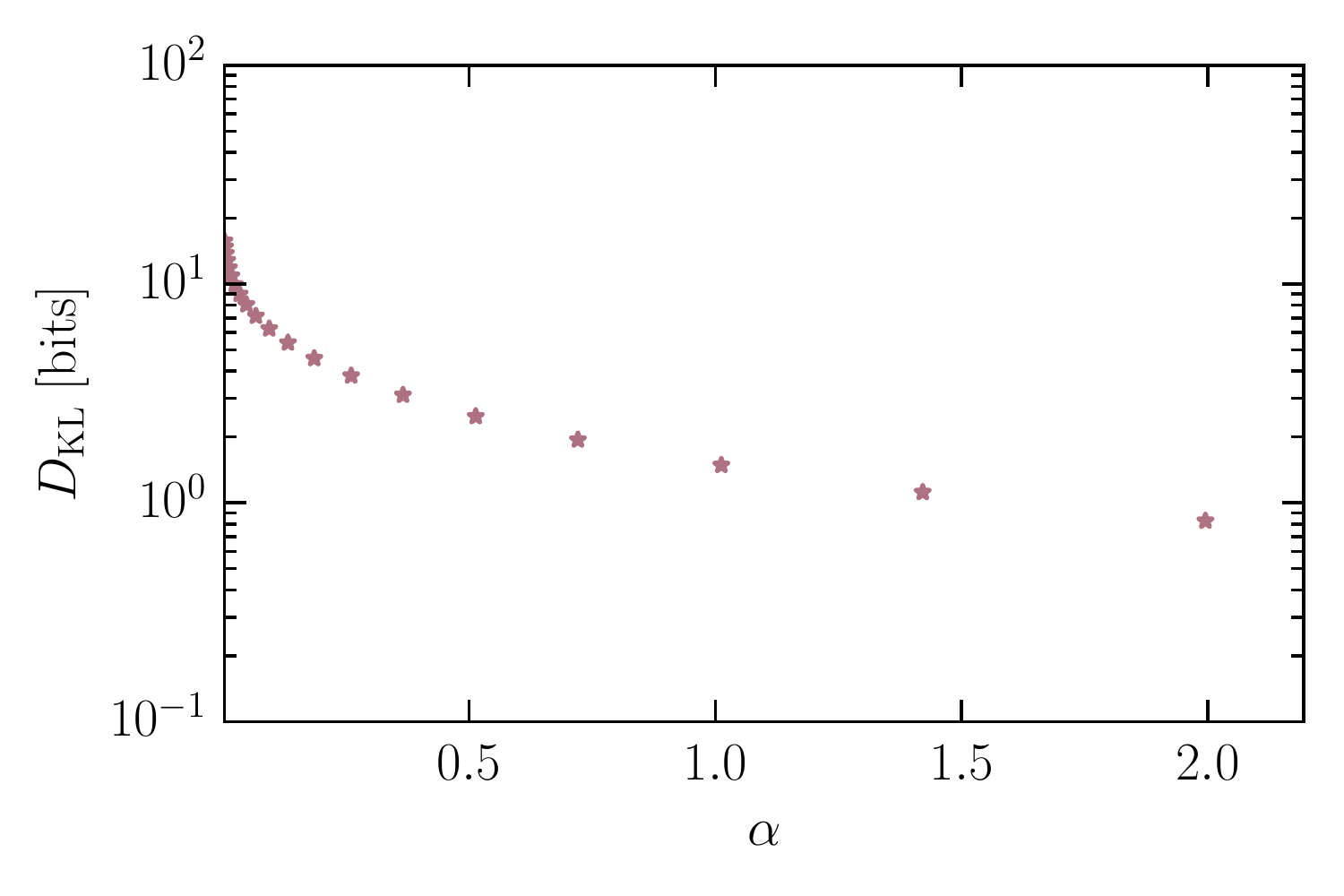}}
\subfigure{\includegraphics[width=0.49\textwidth]{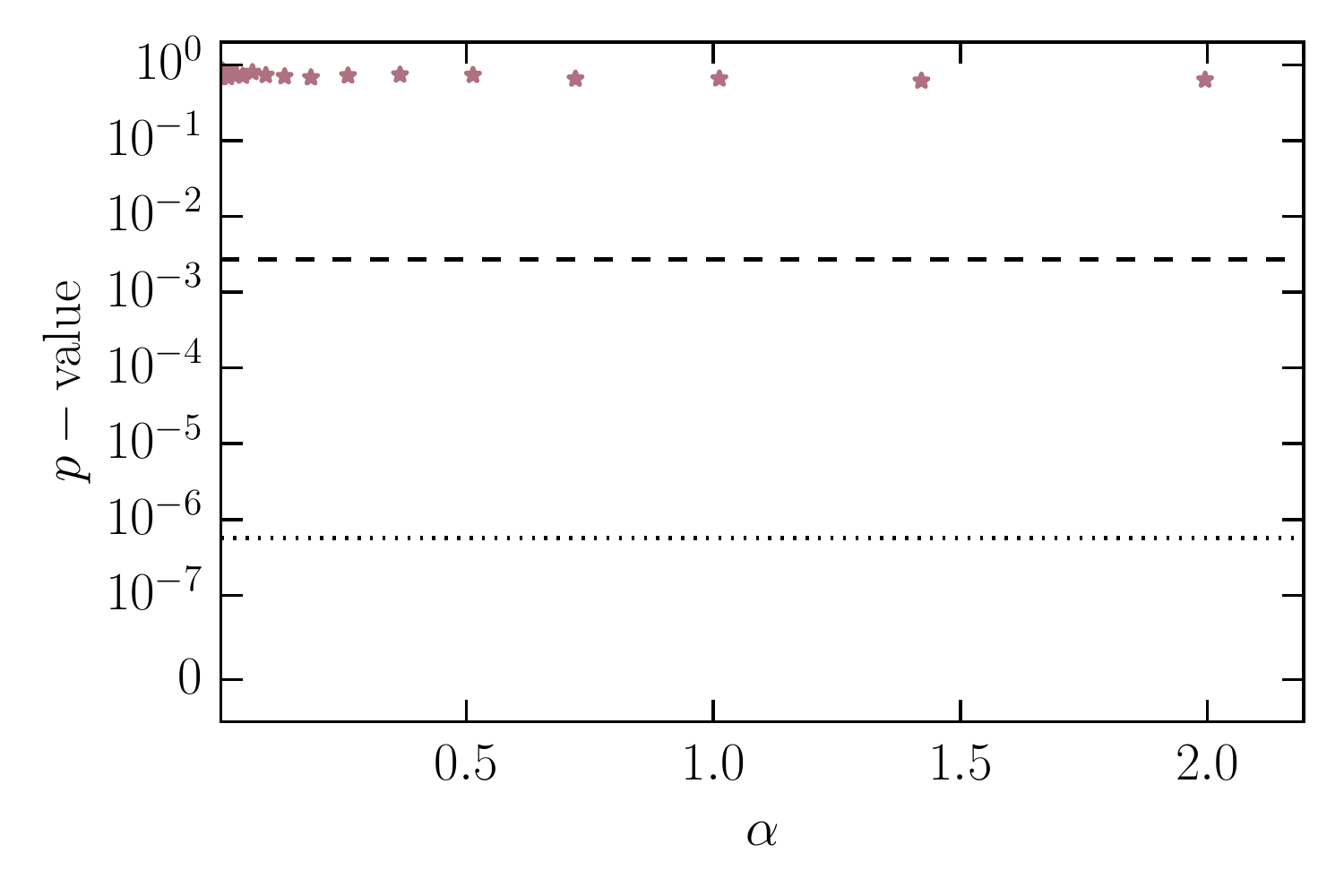}} 
\caption{Illustration of the relative entropy and the $p$-value for the null hypothesis obtained for fitting a simulated data set with $n = 5$ with a polynomial of the same degree as a function of covariance scaling parameter $\alpha$. Results for other combinations of $n, \hat{n}$ are similar. The dashed, horizontal line denotes a $p$-value of $2.7 \times 10^{-3}$, while the dotted line illustrates a $p$-value of $5.7 \times 10^{-7}$. These correspond to the Gaussian equivalent $3 \sigma$ and $5 \sigma$ thresholds.} 
\label{fig:dkl_pval_vs_alpha}
\end{center}
\end{figure}

We further make the assumption that $C_{\textbf{y}^{\prime}} = C_{\textbf{y}}$, which amounts to assuming that the first and second experiment are identical. Intuitively, this assumption is justified as we would like to quantify the consistency of data and model based on our current knowledge. In other words, we would like to understand which models can be ruled out with current observational uncertainties, and thus assuming a repetition of the current observation is a natural choice. Nevertheless, this choice is arbitrary and it is thus instructive to investigate the dependence of our results on the covariance matrix of the second experiment. To this end, we introduce a scaling parameter $\alpha$ that rescales the experimental uncertainties as $C_{\textbf{y}^{\prime}}' = \alpha C_{\textbf{y}^{\prime}}$, i.e., as $\alpha$ approaches zero, the errors of the second experiment decrease. We then compute the relative entropies and $p$-values obtained as a function of $\alpha \in [0, 2]$. In Fig.~\ref{fig:dkl_pval_vs_alpha}, we show the results for fitting the data with a polynomial of degree $\hat{n} = n = 5$, but results for other combinations of polynomials are similar\footnote{We note that we keep all noise realizations constant as we vary $\alpha$ in order to illustrate the variations caused by the change in $\alpha$ alone.}. We see that the relative entropy increases as $\alpha$ decreases, which is expected as the information gain from the model is maximized when measurement uncertainties are minimized. The $p$-value on the other hand stays reasonably constant as we vary $\alpha$. This means that both $D_{\mathrm{KL}}(\textbf{y}, M_{\hat{n}} || \textbf{y})$ and $\langle D_{\mathrm{KL}}(\textbf{y}(M_{\hat{n}}), M_{\hat{n}} || \textbf{y}(M_{\hat{n}})) \rangle$ increase similarly as we make the second experiment more constraining. It therefore appears that the model rejection results are stable for a wide range of $\alpha$ parameters. The only exception, which is not shown in the figure, is that we find an instability as we approach $\alpha = 0$ due to finite machine precision. At $\alpha = 0$, the covariance matrix of the PPD from data and model exhibits large off-diagonal elements and all calculations become numerically unstable, meaning that we cannot compute the information gain for an infinite precision experiment. However, since all experiments in cosmology have a fundamental error floor (e.g. cosmic variance, limited number counts in observable Universe, etc.), this limit is never reached in practice and we can thus safely perform all computations at the fiducial point $C_{\textbf{y}^{\prime}} = C_{\textbf{y}}$.   

\begin{figure}
\begin{center}
\subfigure{\includegraphics[scale=0.45]{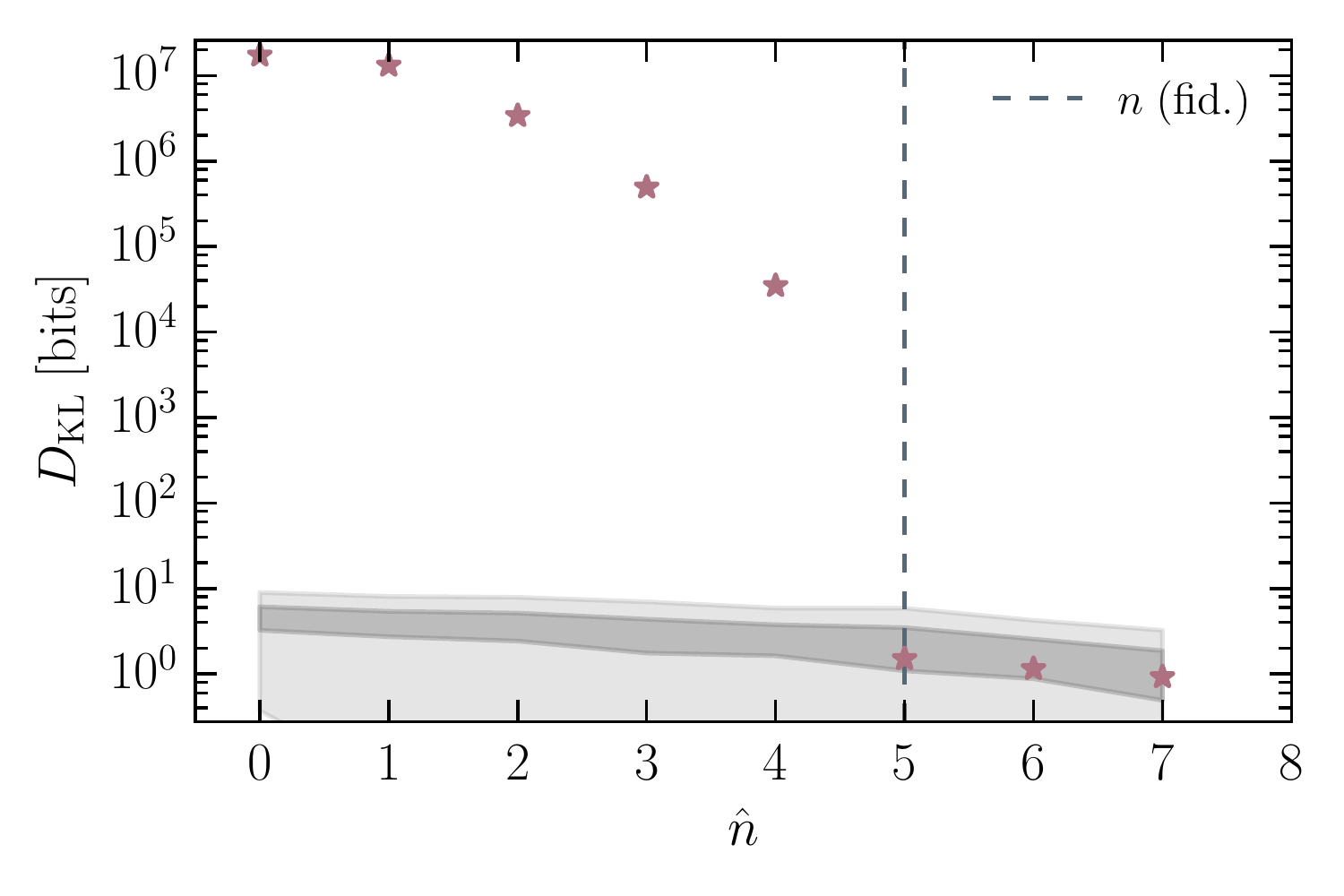}}
\subfigure{\includegraphics[scale=0.45]{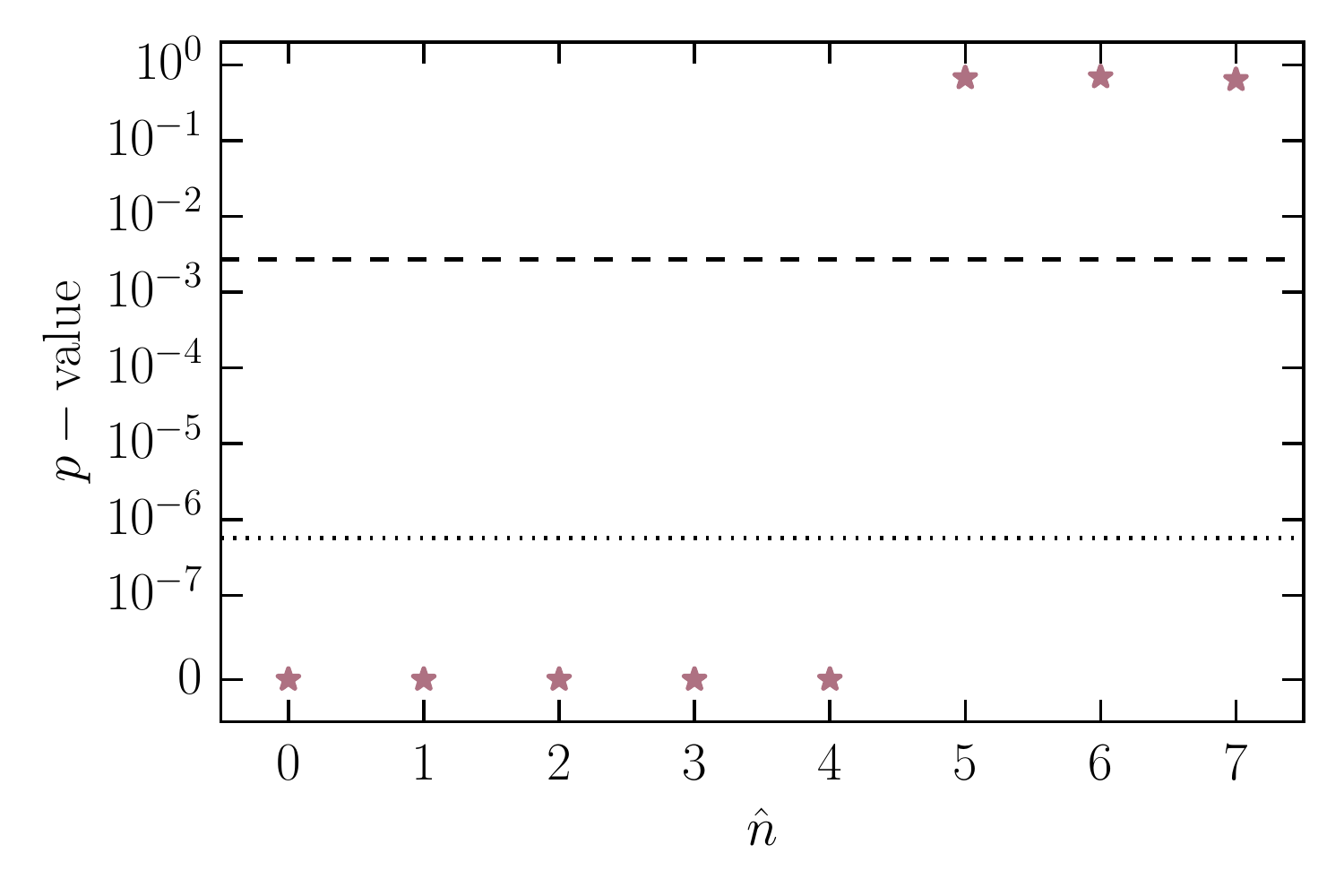}}
\caption{Illustration of $D_{\mathrm{KL}}(\textbf{y}, M_{\hat{n}} || \textbf{y})$ and the $p$-value of $D_{\mathrm{KL}}(\textbf{y}, M_{\hat{n}} || \textbf{y})$ as a function of the degree $\hat{n}$ of the polynomial chosen to fit the data for the toy model described in Sec.~\ref{subsec:toy_model_model_selection}. The gray bands show the $1\sigma$ and $3\sigma$ uncertainties on the information gain from the true model, i.e. $\langle D_{\mathrm{KL}}(\textbf{y}(M_{\hat{n}}), M_{\hat{n}} || \textbf{y}(M_{\hat{n}})) \rangle$. The dashed, horizontal line denotes a $p$-value of $2.7 \times 10^{-3}$, while the dotted line illustrates a $p$-value of $5.7 \times 10^{-7}$. These correspond to the Gaussian equivalent $3 \sigma$ and $5 \sigma$ thresholds.}
\label{fig:dkl_deg}
\end{center}
\end{figure}

In Fig.~\ref{fig:dkl_deg}, we show the results obtained for data generated from a polynomial of fiducial order $n = 5$, which we fit with polynomials of degrees $\hat{n} \in [0, 7]$. The results for all other fiducial polynomials are similar. As can be seen from comparing $D_{\mathrm{KL}}(\textbf{y}, M_{\hat{n}} || \textbf{y})$, $\langle D_{\mathrm{KL}}(\textbf{y}(M_{\hat{n}}), M_{\hat{n}} || \textbf{y}(M_{\hat{n}})) \rangle$ and $\sigma(\langle D_{\mathrm{KL}}(\textbf{y}(M_{\hat{n}}), M_{\hat{n}} || \textbf{y}(M_{\hat{n}})) \rangle)$ in the left panel of Fig.~\ref{fig:dkl_deg}, all polynomials of degrees $\hat{n}$ smaller than the data-generating degree $n$ are ruled out with high significance as they result in relative entropies significantly larger than expected for the true model. All models with $\hat{n} \geq n$ on the other hand, give relative entropies consistent with expectations from the true model, suggesting that these models provide a good fit to the data, as expected. Finally, in the right panel of Fig.~\ref{fig:dkl_deg} we show the $p$-values corresponding to these relative entropies and we can see that the $p$-values reflect the results discussed for $D_{\mathrm{KL}}(\textbf{y}, M_{\hat{n}} || \textbf{y})$. 

\subsection{Application to cosmological data}\label{subsec:cosmology_applic}

We test the method described above on cosmological data by applying it to the SNe Ia sample from the Joint Lightcurve Analysis (JLA) \cite{Betoule:2014}, which comprises data from SDSS-II \cite{Frieman:2008, Kessler:2009, Sollerman:2009, Lampeitl:2010, Campbell:2013}, the Supernova Legacy Survey (SNLS) \cite{Astier:2006, Sullivan:2011}, the HST \cite{Riess:2007, Suzuki:2012} and additional low-redshift experiments \cite{Betoule:2014}.\footnote{The data can be found at: $\tt{http://supernovae.in2p3.fr/sdss\_snls\_jla/ReadMe.html}$.} The JLA data consists of light curve parameters for $N_{\mathrm{SNe}} = 740$ SNe Ia, which can be used to calculate distance moduli $\mu(z_{\mathrm{SNe}})$, as well as their covariance. 

The distance modulus of an SNe Ia at redshift $z_{\mathrm{SNe}}$ is given by 
\begin{equation}
\mu(z_{\mathrm{SNe}}) = 5 \log_{10}\left(\frac{d_{\mathrm{L}}(z_{\mathrm{SNe}})}{10 [\mathrm{pc}]}\right),
\label{eq:distmod}
\end{equation} 
where $d_{\mathrm{L}}(z_{\mathrm{SNe}})$ denotes the luminosity distance. The observed supernovae absolute peak magnitudes and thus their distance moduli have been found to depend on several SNe Ia and host galaxy properties. We therefore follow Ref.~\cite{Betoule:2014} and parametrize the observed distance moduli as
\begin{equation}
\mu_{\mathrm{obs}} = m_{\mathrm{B}}^{*} - (M_{\mathrm{B}} - \alpha X_{1} + \beta C),
\end{equation}
where $m_{\mathrm{B}}^{*}$ denotes the observed peak magnitude of the SNe in rest frame $\tt{B}$-band, $M_{\mathrm{B}}$ is the absolute magnitude, $C$ is the color of the SNe and $X_{1}$ is the so-called stretch parameter, which quantifies the duration of the SNe explosion. The parameters $\alpha, \beta$ and $M_{\mathrm{B}}$ are nuisance parameters and both $M_{\mathrm{B}}$ and $\beta$ were found to depend on supernova host galaxy properties \cite{Sullivan:2010, Johansson:2013}. In order to take these effects into account, we follow Ref.~\cite{Betoule:2014} and set 
\begin{equation}
M_{\mathrm{B}} = 
    \begin{cases}
      M^{1}_{\mathrm{B}} & \text{if } M_{\text{stellar}} < 10^{10} M_{\odot}, \\
      M^{1}_{\mathrm{B}} + \Delta M       & \text{otherwise},
    \end{cases}
\end{equation}
where $\Delta M$ is a nuisance parameter additional to $\alpha, \beta$ and $M^{1}_{\mathrm{B}}$.

\begin{table*}
\caption{Summary of the cosmological models considered in this work.} \label{tab:models}
\begin{center}
\begin{tabular}{ccc>{\centering}m{2.5cm}>{\centering}m{2.8cm}@{}m{0pt}@{}}
\hline\hline 
Model no. & Cosmology & Density constraints & Parameters w/o CMB & Parameters w/ CMB & \\ \hline      
0 & CDM & $\Omega_{\Lambda} = 0, \Omega_{\mathrm{K}} = 0$ & $\alpha, \allowbreak \beta, \allowbreak M^{1}_{\mathrm{B}}, \allowbreak \Delta M$ & $h, \allowbreak \Omega_{\mathrm{b}}, \allowbreak \alpha, \allowbreak \beta, \allowbreak M^{1}_{\mathrm{B}}, \allowbreak \Delta M$ & \\
1 & curved CDM & $\Omega_{\Lambda} = 0$ & $\Omega_{\mathrm{m}}, \allowbreak \alpha, \allowbreak \beta, \allowbreak M^{1}_{\mathrm{B}}, \allowbreak \Delta M$ & $h, \allowbreak \Omega_{\mathrm{m}}, \allowbreak \Omega_{\mathrm{b}}, \allowbreak \alpha, \allowbreak \beta, \allowbreak M^{1}_{\mathrm{B}}, \allowbreak \Delta M$ & \\
2 & $\Lambda$CDM & $\Omega_{\mathrm{K}} = 0$ & $\Omega_{\mathrm{m}}, \allowbreak \alpha, \allowbreak \beta, \allowbreak M^{1}_{\mathrm{B}}, \allowbreak \Delta M$ & $h, \allowbreak \Omega_{\mathrm{m}}, \allowbreak \Omega_{\mathrm{b}}, \allowbreak \alpha, \allowbreak \beta, \allowbreak M^{1}_{\mathrm{B}}, \allowbreak \Delta M$ & \\
3 & curved $\Lambda$CDM & - & $\Omega_{\mathrm{m}}, \allowbreak \Omega_{\Lambda}, \allowbreak \alpha, \allowbreak \beta, \allowbreak M^{1}_{\mathrm{B}}, \allowbreak \Delta M$ & $h, \allowbreak \Omega_{\mathrm{m}}, \allowbreak \Omega_{\Lambda}, \allowbreak \Omega_{\mathrm{b}}, \allowbreak \alpha, \allowbreak \beta, \allowbreak M^{1}_{\mathrm{B}}, \allowbreak \Delta M$ & \\
4 & $w_{0}$CDM & $\Omega_{\mathrm{K}} = 0$ & $\Omega_{\mathrm{m}}, \allowbreak w_{0}, \allowbreak \alpha, \allowbreak \beta, \allowbreak M^{1}_{\mathrm{B}}, \allowbreak \Delta M$ & $h, \allowbreak \Omega_{\mathrm{m}}, \allowbreak w_{0}, \allowbreak \Omega_{\mathrm{b}}, \allowbreak \alpha, \allowbreak \beta, \allowbreak M^{1}_{\mathrm{B}}, \allowbreak \Delta M$ & \\
5 & $w_{0}w_{a}$CDM & $\Omega_{\mathrm{K}} = 0$ & $\Omega_{\mathrm{m}}, \allowbreak w_{0}, \allowbreak w_{a}, \allowbreak \alpha, \allowbreak \beta, M^{1}_{\mathrm{B}}, \allowbreak \Delta M$ & $h, \allowbreak \Omega_{\mathrm{m}}, \allowbreak w_{0}, \allowbreak w_{a}, \allowbreak \Omega_{\mathrm{b}}, \allowbreak \alpha, \allowbreak \beta, \allowbreak M^{1}_{\mathrm{B}}, \allowbreak \Delta M$ & \\
 \hline \hline
\end{tabular}
\end{center}
\end{table*}

We apply the model rejection method described above to quantifying the consistency between the JLA data and several cosmological models. In this analysis, we consider six models: (i) CDM, (ii) curved CDM, (iii) $\Lambda$CDM, (iv) curved $\Lambda$CDM, (v) $w_{0}$CDM and (vi) $w_{0}w_{a}$CDM. For each model, we first compute posterior distributions for all model parameters (cosmological and nuisance) in a Monte Carlo Markov Chain (MCMC) using the publicly-available code \texttt{CosmoHammer}\footnote{\texttt{CosmoHammer} is based on \texttt{emcee} \cite{Foreman-Mackey:2013} and the code can be found at: \texttt{http://cosmo-docs.phys.ethz.ch/cosmoHammer.}} \cite{Akeret:2013}. The parameters varied for each considered model are given in Tab.~\ref{tab:models}, where $h$ is the dimensionless Hubble parameter, $\Omega_{\mathrm{m}}$ is the fractional matter density today, $\Omega_{\mathrm{b}}$ is the fractional baryon density today, $\Omega_{\mathrm{K}}$ is the fractional curvature density today, $\Omega_{\Lambda}$ is the fractional dark energy density today and $w_{0}, w_{a}$ parametrize the dark energy equation of state parameter as $w(a) = w_{0} + w_{a}(1-a)$ \cite{Chevallier:2001, Linder:2003}. All theoretical predictions are computed using \texttt{PyCosmo} \cite{Refregier:2017} and we give the posterior means derived for each cosmological model in Appendix \ref{ap:posteriors}. We use these posteriors and the SNe Ia likelihood given in Ref.~\cite{Betoule:2014} for computing the relative entropies defined above. Following the discussion in Sec.~\ref{subsec:toy_model_model_selection}, we assume repetition of an identical experiment when computing all PPDs. In order to reduce the dimensionality of the PPDs, we bin the SNe Ia data into 30 equally log-spaced bins in the redshift range $0.01 < z < 1.3$. We then follow Sec.~\ref{subsec:implementation} and compute the observed relative entropy $D_{\mathrm{KL}}(\boldsymbol{\mu}, M_{i} || \boldsymbol{\mu})$, the expected relative entropy $\langle D_{\mathrm{KL}}(\boldsymbol{\mu}(M_{i}), M_{i} || \boldsymbol{\mu}(M_{i})) \rangle$ and the standard deviation of the expected relative entropy $\sigma(\langle D_{\mathrm{KL}}(\boldsymbol{\mu}(M_{i}), M_{i} || \boldsymbol{\mu}(M_{i})) \rangle)$ for all models $M_{i}, i = 0,\ldots, 5$ considered. However, the algorithm for computing $\langle D_{\mathrm{KL}}(\boldsymbol{\mu}(M_{i}), M_{i} || \boldsymbol{\mu}(M_{i})) \rangle$ and $\sigma(\langle D_{\mathrm{KL}}(\boldsymbol{\mu}(M_{i}), M_{i} || \boldsymbol{\mu}(M_{i})) \rangle)$ requires us to determine the best-fit model parameter values for each simulated data set, which is prohibitively expensive for the present case. Therefore, we resort to an alternative method that avoids this step: instead of first sampling a realization of the data distribution $p(\boldsymbol{\mu} \vert \boldsymbol{\mu}_{\mathrm{t}}, C_{\boldsymbol{\mu}_{\mathrm{t}}})$, and then determining the posterior distribution of the parameters, we first sample a realization of model parameters from the posterior distribution determined from the observed data, $p(\boldsymbol{\theta} \vert \boldsymbol{\mu}, C_{\boldsymbol{\mu}})$, and then use these values to compute a corresponding data vector. These two approaches are equivalent, provided we account for residual noise when generating data from model parameters. In this work, we assume a simple error model and set the covariance of the residuals to $C_{\mathrm{res}} =  \sfrac{(N_{\mathrm{SNe}} - N_{p})}{N_{\mathrm{SNe}}} C_{\boldsymbol{\mu}}$, i.e. we account for the reduction in the covariance caused by using the data to fit $N_{p}$ model parameters. Once we have created corresponding model parameter and data realizations, we proceed to compute $\langle D_{\mathrm{KL}}(\boldsymbol{\mu}(M_{i}), M_{i} || \boldsymbol{\mu}(M_{i})) \rangle$ and $\sigma(\langle D_{\mathrm{KL}}(\boldsymbol{\mu}(M_{i}), M_{i} || \boldsymbol{\mu}(M_{i})) \rangle)$ as described in Sec.~\ref{subsec:implementation}, using $N = 1000$ samples of length $\bar{N} = 1000$ each. We test our modeling of residual errors by running a smaller number of realizations, $N = 200$, in which we determine the best-fit parameters for each simulated data set using a Particle Swarm Optimizer, as implemented in \texttt{CosmoHammer}. This analysis yields results equivalent to those obtained from the approximate method and we therefore resort to the latter as it is computationally less expensive. For a more detailed discussion of this test, the reader is referred to Appendix \ref{ap:test-sampling}.

\begin{figure}
\begin{center}
\subfigure{\includegraphics[width=0.49\textwidth]{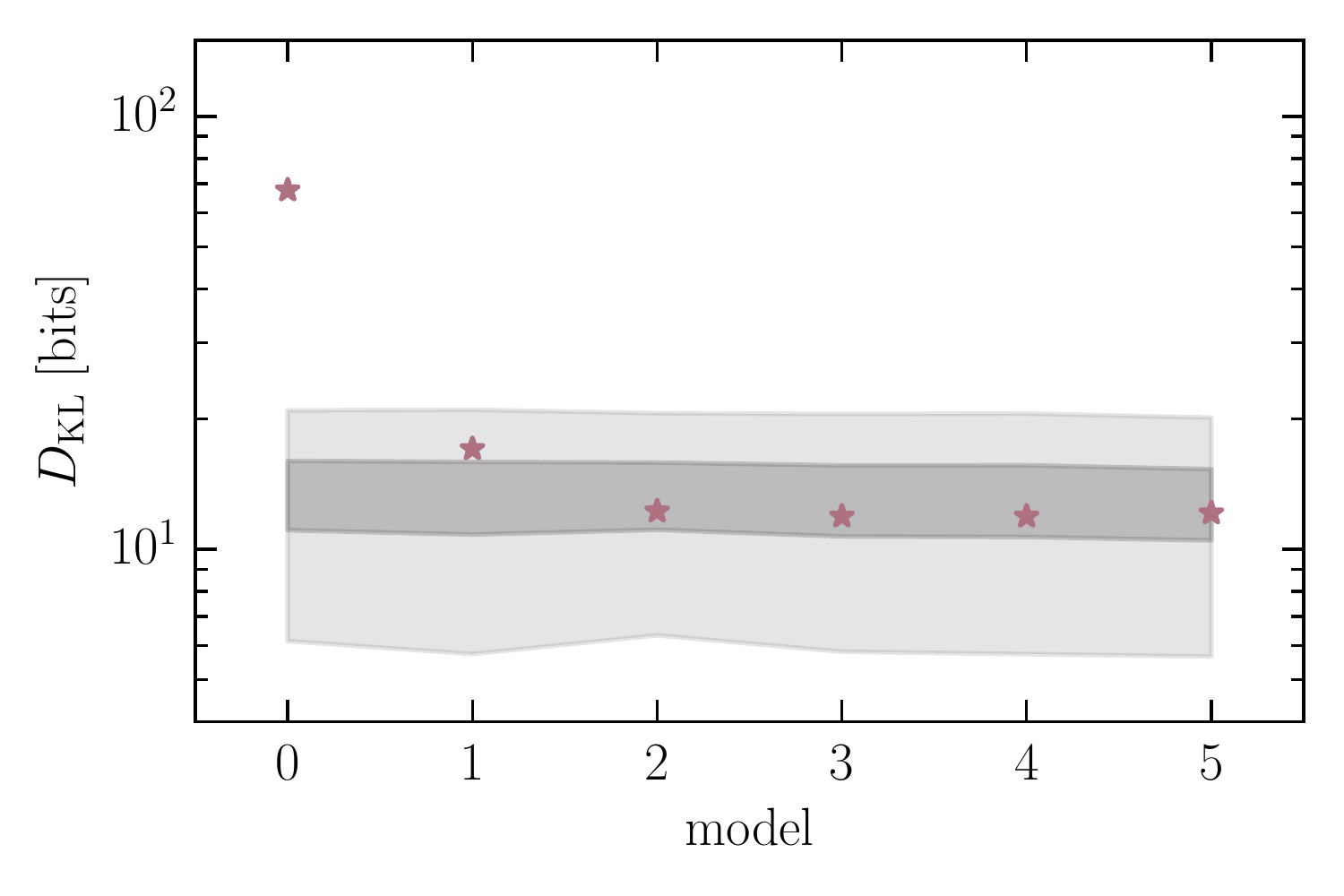}}
\subfigure{\includegraphics[width=0.49\textwidth]{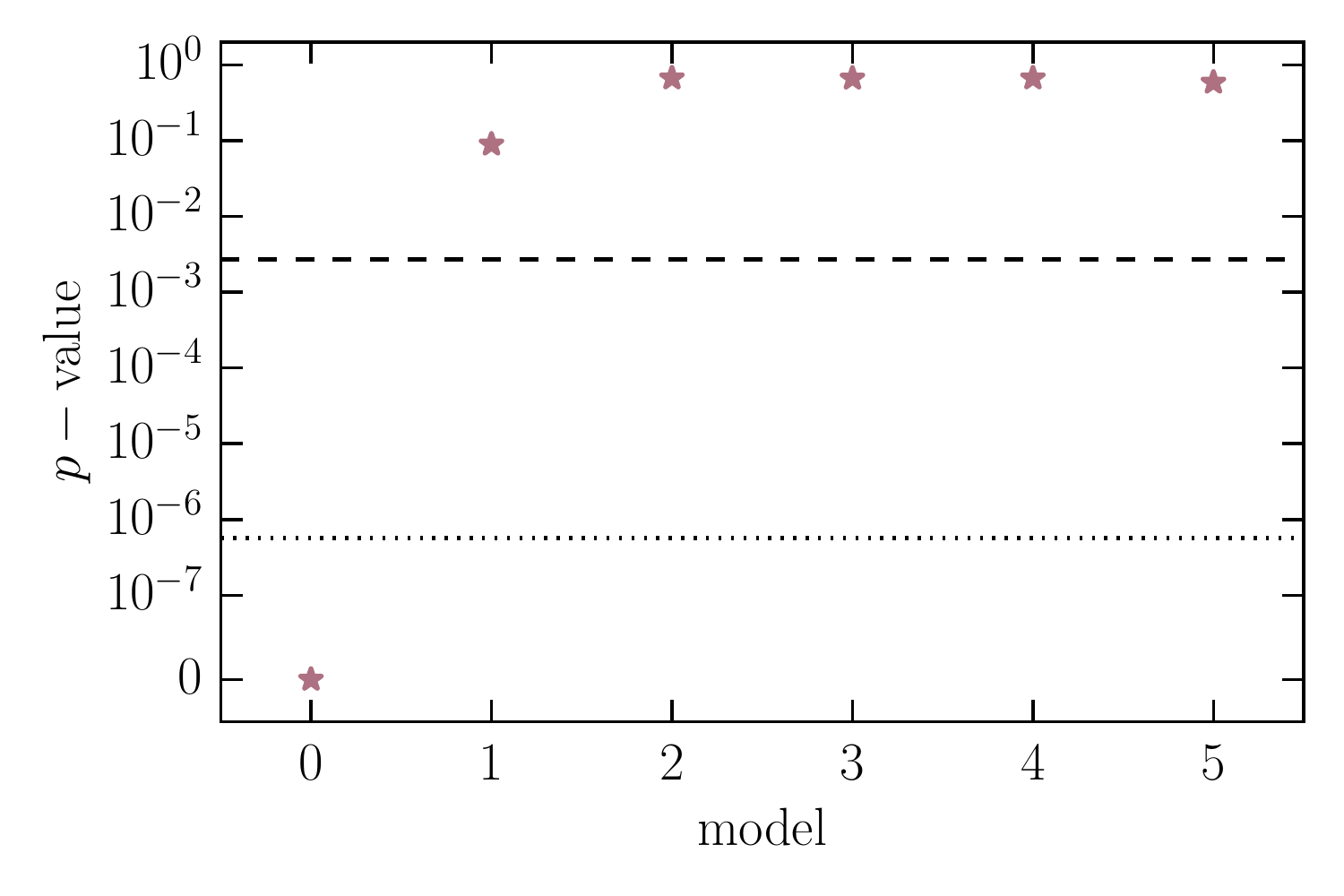}} 
\caption{Illustration of $D_{\mathrm{KL}}(\boldsymbol{\mu}, M_{i} || \boldsymbol{\mu})$ and the $p$-value of $D_{\mathrm{KL}}(\boldsymbol{\mu}, M_{i} || \boldsymbol{\mu})$ obtained for a fit to JLA SNe Ia as a function of the cosmological model chosen to fit the data. The gray bands show the $1\sigma$ and $3\sigma$ errors on the information gain from the true model, i.e. $\langle D_{\mathrm{KL}}(\boldsymbol{\mu}(M_{i}), M_{i} || \boldsymbol{\mu}(M_{i})) \rangle$.  The dashed, horizontal line denotes a $p$-value of $2.7 \times 10^{-3}$, while the dotted line illustrates a $p$-value of $5.7 \times 10^{-7}$. These correspond to the Gaussian equivalent $3 \sigma$ and $5 \sigma$ thresholds. The correspondence between model number and cosmological model is given in Tab.~\ref{tab:models}.} 
\label{fig:dkl_vs_model_jla}
\end{center}
\end{figure}

The relative entropies and associated $p$-values for each cosmological model are shown in Fig.~\ref{fig:dkl_vs_model_jla}. As we can see, the CDM model is clearly ruled out by JLA supernovae data with a $p$-value $p < 5.7 \times 10^{-7}$, whereas curved CDM cannot be ruled out using supernovae data alone, as it provides an acceptable fit with $p$-value $p = 0.09$. This is similar to the results found in Refs.~\cite{Nielsen:2016, Matthews:2017}, albeit at a different significance, which is probably due to the different methodologies employed. Finally, the $\Lambda$CDM, curved $\Lambda$CDM, $w_{0}$CDM and $w_{0}w_{a}$CDM models are all consistent with the data, as the probabilities to obtain a relative entropy as large or larger than the one observed equal at least 0.6. In order to test the stability of our results to the choice of binning scheme, we repeat this analysis for several cases: (i) 30 bins with constant number of SNe Ia, and (ii) $n_{\mathrm{bin}} = 65, 100, 200, 300$ equally log-spaced bins. For all cases, we find changes in the numerical values of both the relative entropies and the $p$-values, but our conclusions remain unchanged. In general, we see a trend of increasing $p$-values for larger data size. For example, when using the unbinned SNe likelihood to compute the KL divergences, we find that CDM is excluded at the $4\sigma$ level. This is not entirely unexpected, as $\chi^{2}$ analyses are prone to binning dependence. We therefore conclude that the proposed method suffers somewhat from these instabilities but yields stable conclusions for a wide range of binning schemes\footnote{We note that we have investigated the Gaussianity of the SNe Ia data as a function of binning scheme, finding that Gaussianity cannot be excluded for all binning schemes considered. Therefore, failure of the Gaussian assumption cannot explain the binning sensitivities seen.}.

The potential degeneracy between curvature and cosmological constant, exhibited by SNe Ia data, can be broken by combining the SNe constraints with external data. In this work, we complement the SNe data with constraints on $\Omega_{\mathrm{K}}$ in form of the CMB shift parameter $R$, which is defined as
\begin{equation}
R = \sqrt{\Omega_{\mathrm{m}} H_{0}^{2}}\frac{D_{\mathrm{A}}(z_{\star})}{c},
\end{equation}
where $D_{\mathrm{A}}(z_{\star})$ denotes the comoving angular diameter distance to the redshift of decoupling $z_{\star}$, $H_{0}$ is the Hubble parameter and $c$ denotes the speed of light. We include the constraint on $R$ obtained by the Planck Collaboration in their second data release \cite{Planck-Collaboration:2016ai}. With this added constraint, we repeat the above analysis, additionally varying the dimensionless Hubble parameter $h$ and the fractional baryon density $\Omega_{\mathrm{b}}$ (see Tab.~\ref{tab:models}). As we only include one additional data point for the CMB, we do not distinguish between residual and measurement uncertainty as we did for the SNe sample.  

\begin{figure}
\begin{center}
\subfigure{\includegraphics[width=0.49\textwidth]{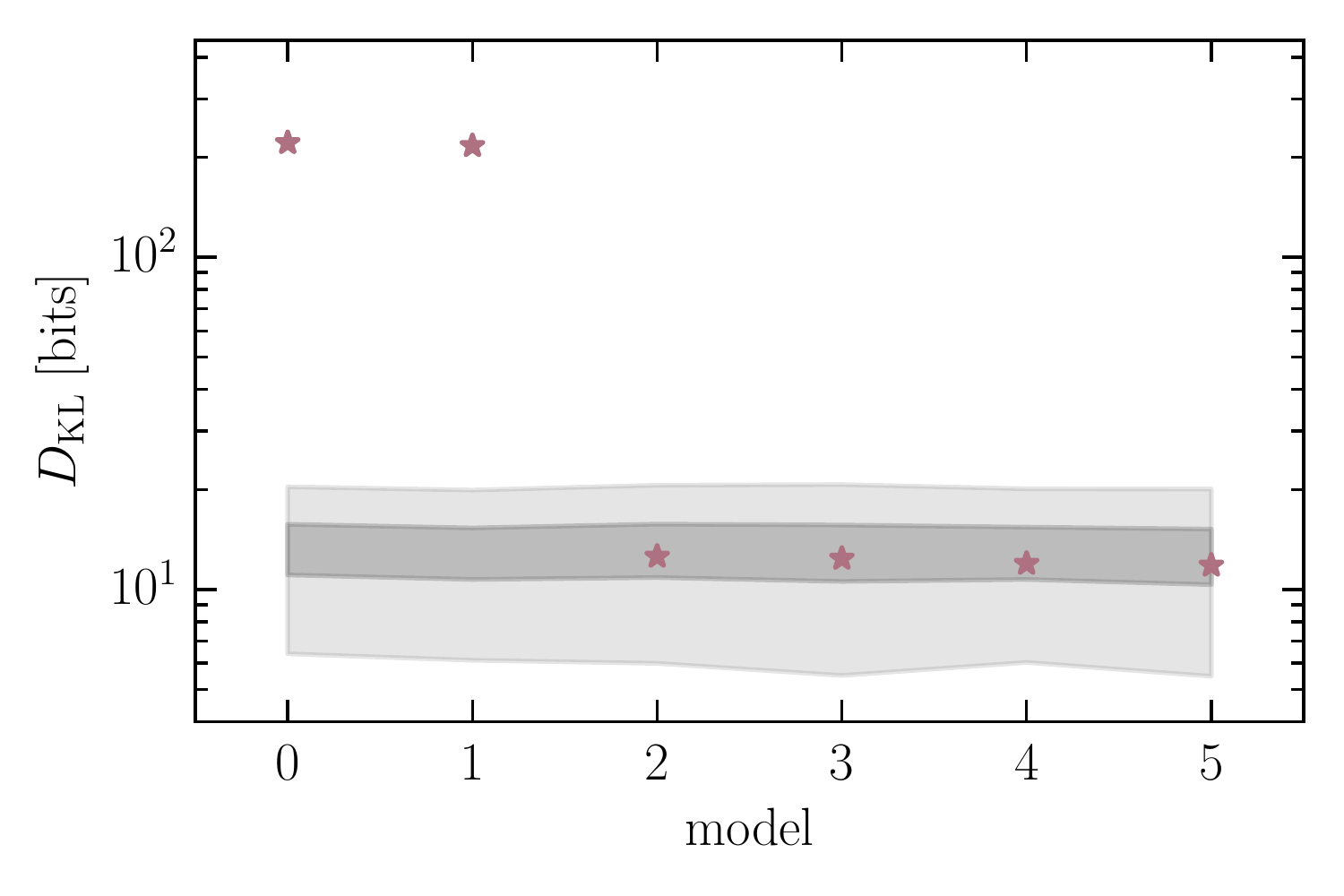}}
\subfigure{\includegraphics[width=0.49\textwidth]{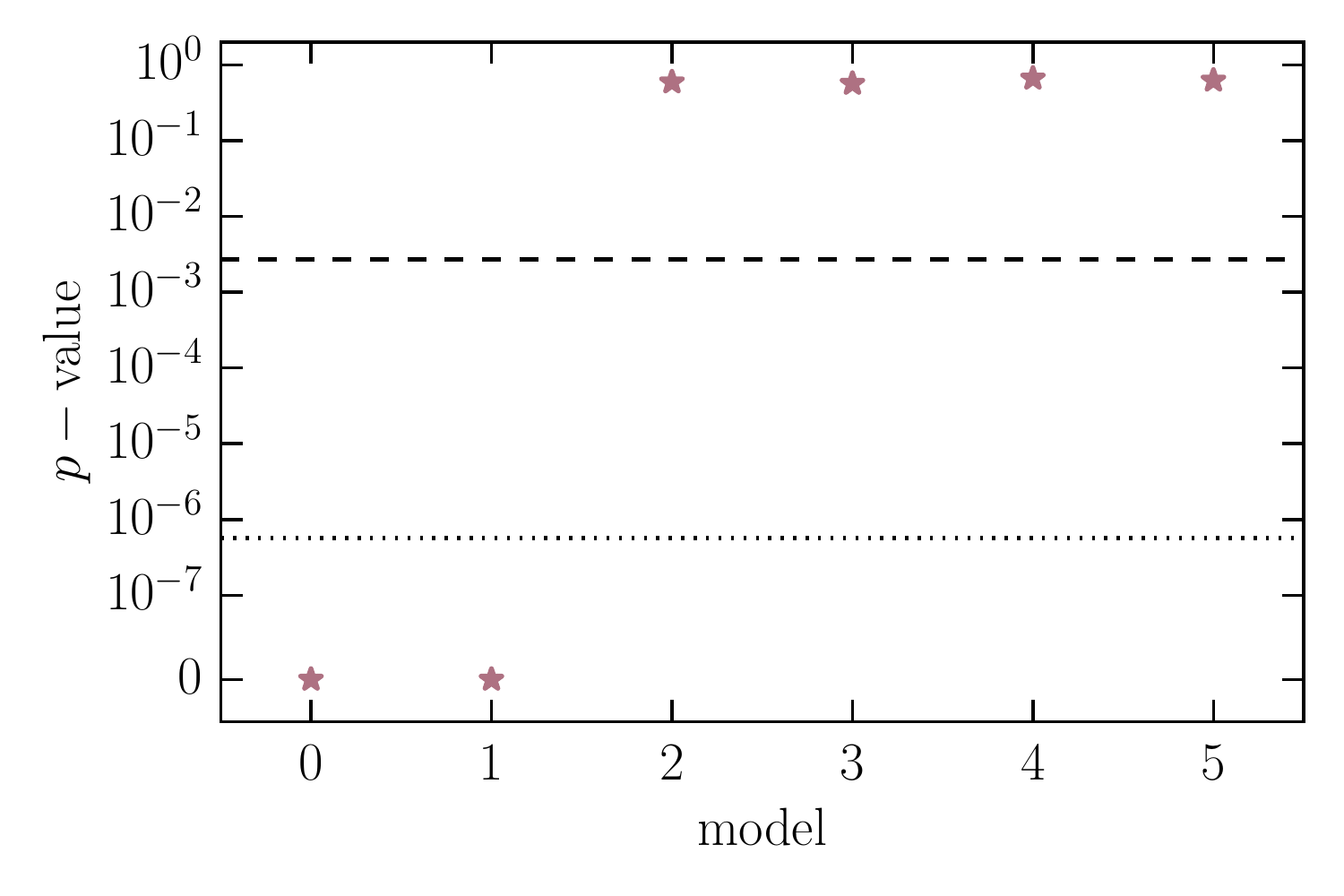}} 
\caption{Illustration of $D_{\mathrm{KL}}(\boldsymbol{\mu}, M_{i} || \boldsymbol{\mu})$ and the $p$-value of $D_{\mathrm{KL}}(\boldsymbol{\mu}, M_{i} || \boldsymbol{\mu})$ obtained for a fit to JLA SNe Ia and the CMB shift parameter $R$ as a function of the cosmological model chosen to fit the data. The gray bands show the $1\sigma$ and $3\sigma$ errors on the information gain from the true model, i.e. $\langle D_{\mathrm{KL}}(\boldsymbol{\mu}(M_{i}), M_{i} || \boldsymbol{\mu}(M_{i})) \rangle$. The dashed, horizontal line denotes a $p$-value of $2.7 \times 10^{-3}$, while the dotted line illustrates a $p$-value of $5.7 \times 10^{-7}$. These correspond to the Gaussian equivalent $3 \sigma$ and $5 \sigma$ thresholds. The correspondence between model number and cosmological model is given in Tab.~\ref{tab:models}.} 
\label{fig:dkl_vs_model_jla+cmb}
\end{center}
\end{figure}

Fig.~\ref{fig:dkl_vs_model_jla+cmb} shows the obtained relative entropies and associated $p$-values for the cosmological models considered. We see that the inclusion of the CMB constraint has the effect that besides CDM, curved CDM is also ruled out with very high significance. The remaining models, i.e. $\Lambda$CDM, curved $\Lambda$CDM, $w_{0}$CDM and $w_{0}w_{a}$CDM, still provide good fits to the data. This shows that SNe Ia data combined with a minimal constraint on the matter density and the curvature of the Universe are able to clearly rule out both non-accelerating cosmological models considered in this work. All accelerating models, i.e. $\Lambda$CDM and its extensions, on the other hand, are consistent with the data. 

We again compare different binning schemes and find our results to be unaffected in all the considered cases (even in the unbinned case). This is probably due to the fact that the disagreement between the non-accelerated models and the data is so clear that it is insensitive to any analysis choice.

These results are consistent with earlier works that have also found SNe Ia data combined with CMB or BAOs to rule out non-accelerated models to a very high significance (e.g. \cite{Rubin:2016, Haridasu:2017, Nielsen:2016, Matthews:2017}). The above analysis thus demonstrates the applicability of the methodology described in Section \ref{subsec:implementation} to cosmological data sets.

\subsubsection{Comparison to Bayesian evidence}

In order to understand if the low significance with which SNe Ia data alone exclude curved CDM is a consequence of the model rejection framework proposed here, we perform a similar analysis using the Bayesian evidence, which is a popular tool for model comparison (see e.g. \cite{Knuth:2015}). The evidence $p(\textbf{y}|M)$ is the normalization in Bayes' theorem and gives the probability of obtaining data $\textbf{y}$ given a model $M$, i.e.
\begin{equation}
p(\textbf{y}|M) = \int \mathrm{d}\boldsymbol{\theta} \; p(\boldsymbol{\theta}) p(\textbf{y}|\boldsymbol{\theta}),
\end{equation}
where $p(\boldsymbol{\theta})$ and $p(\textbf{y}|\boldsymbol{\theta})$ denote the prior and the likelihood respectively.

In order to perform model rejection with the Bayesian evidence, we employ a procedure analogous to the one used for the relative entropy: we test the goodness of fit of any model by comparing the observed evidence to its expectation value under the null hypothesis. The expectation value of the evidence is given by
\begin{equation}
\langle p(\textbf{y}|M) \rangle = \int \mathrm{d}\textbf{y} \; p(\textbf{y}|M) p(\textbf{y}|M) = \int \mathrm{d}\textbf{y} \; p(\textbf{y}|M) \int \mathrm{d}\boldsymbol{\theta} \; p(\boldsymbol{\theta}) p(\textbf{y}|\boldsymbol{\theta}).
\end{equation} 
In this work, we compute $\langle p(\textbf{y}|M) \rangle$ through Monte Carlo integration, i.e. we first sample a set of model parameters $\boldsymbol{\theta}$ from the prior distribution $p(\boldsymbol{\theta})$, then we sample a data realization from $p(\textbf{y}|\boldsymbol{\theta})$ and compute the corresponding evidence $p(\textbf{y}|M)$. We repeat this process $N=100$ times and average over the obtained values in order to estimate $\langle p(\textbf{y}|M) \rangle$ and $\sigma(\langle p(\textbf{y}|M) \rangle)$. For all evidence calculations, we employ the publicly available code \texttt{MCEvidence}\footnote{The code is available at: \texttt{https://github.com/yabebalFantaye/MCEvidence}.} \cite{Heavens:2017a, Heavens:2017}, which computes the evidence directly from Monte Carlo Markov Chains.

\begin{figure}
\begin{center}
\subfigure{\includegraphics[width=0.49\textwidth]{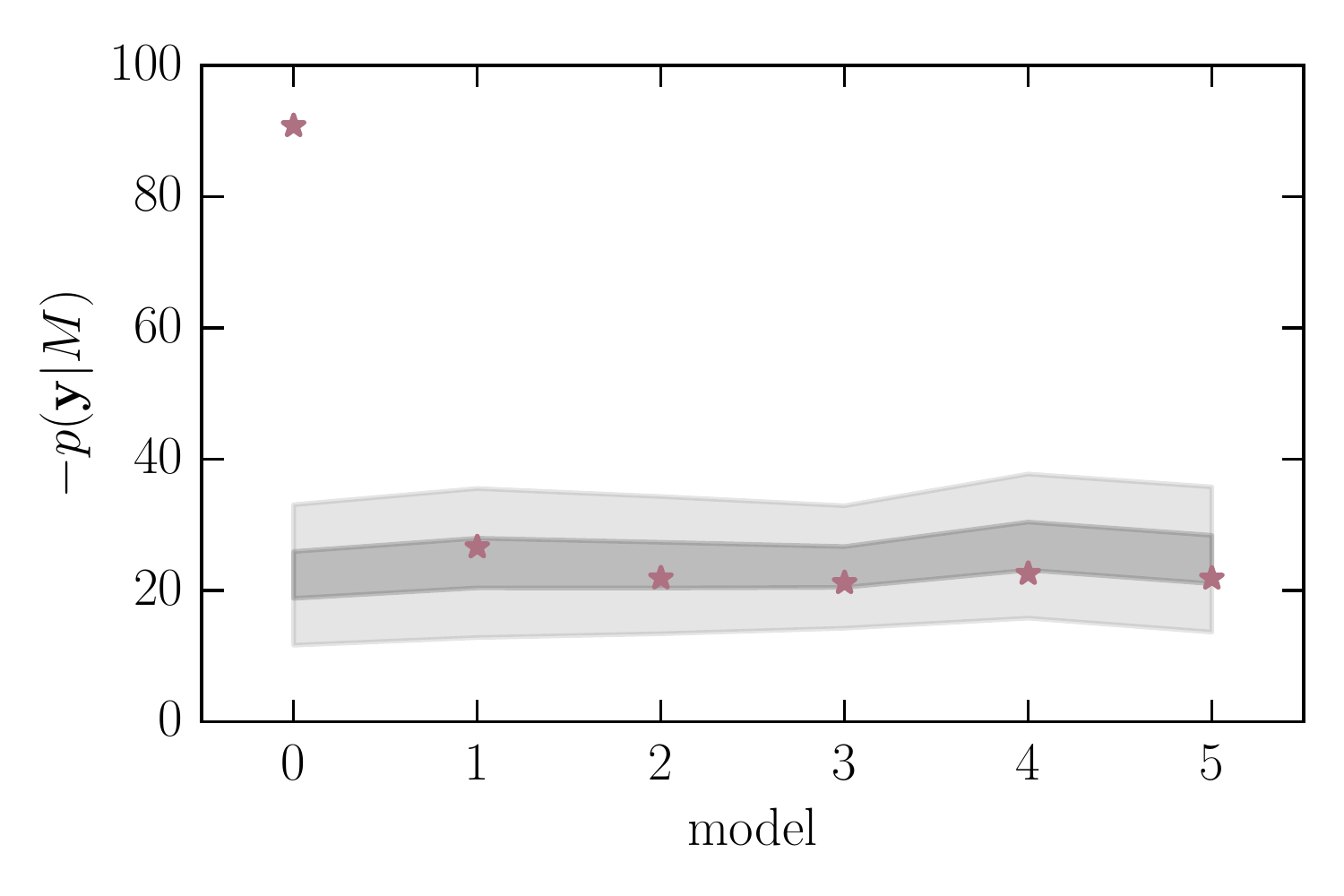}}
\subfigure{\includegraphics[width=0.49\textwidth]{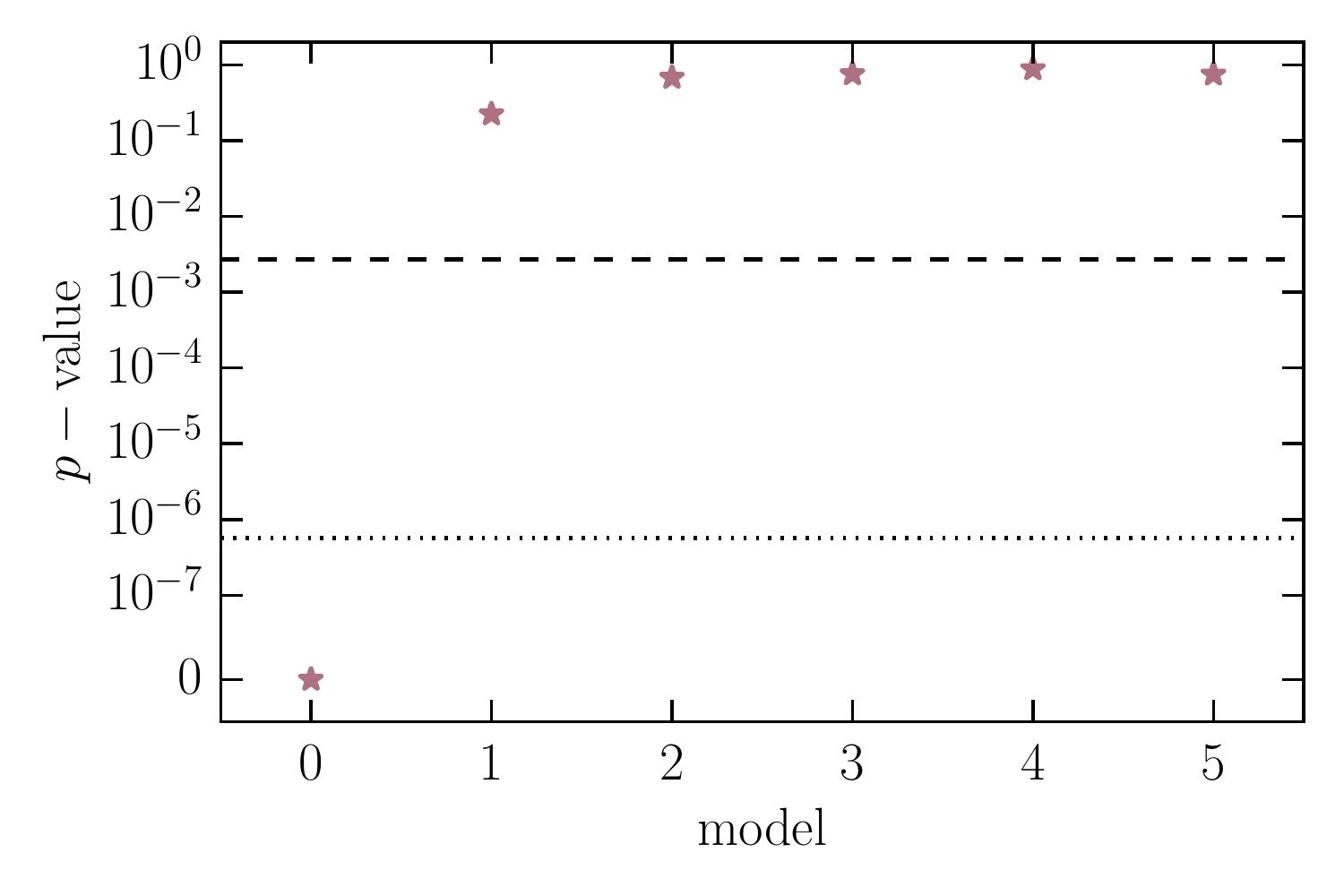}} 
\caption{Illustration of the evidence $p(\textbf{y}|M)$ and the $p$-value of $p(\textbf{y}|M)$ obtained for a fit to JLA SNe Ia as a function of the cosmological model chosen to fit the data. The gray bands show the $1\sigma$ and $3\sigma$ errors on the expected evidence, i.e. $\langle p(\textbf{y}|M) \rangle$. The dashed, horizontal line denotes a $p$-value of $2.7 \times 10^{-3}$, while the dotted line illustrates a $p$-value of $5.7 \times 10^{-7}$. These correspond to the Gaussian equivalent $3 \sigma$ and $5 \sigma$ thresholds. The correspondence between model number and cosmological model is given in Tab.~\ref{tab:models}.} 
\label{fig:evidence_vs_model_jla}
\end{center}
\end{figure}

Fig.~\ref{fig:evidence_vs_model_jla} shows the results obtained when fitting the cosmological models given in Tab.~\ref{tab:models} to SNe Ia data alone, which we have binned into 30 equally log-spaced bins in the redshift range $0.01 < z < 1.3$. These results are similar to those for the relative entropy, shown in Fig.~\ref{fig:dkl_vs_model_jla}. As before, comparing the observed to the expected Bayesian evidence, we find that SNe Ia data clearly rule out CDM, while curved CDM and all accelerated models considered are allowed. Finally, we perform the analogous analysis using unbinned SNe Ia data and find similar results, albeit at generally higher $p$-values, similar to the results found using the method based on the relative entropy.

Typically, the evidence is applied to model comparison rather than model rejection: two competing models, $M_{1}$ and $M_{2}$, are compared using the ratio of their posterior probabilities $p(M_{1}|\textbf{y})$, $p(M_{2}|\textbf{y})$
\begin{equation}
\frac{p(M_{1}|\textbf{y})}{p(M_{2}|\textbf{y})} = \frac{p(M_{1}) \; p(\textbf{y}|M_{1})}{p(M_{2}) \; p(\textbf{y}|M_{2})} =  \frac{p(M_{1})}{p(M_{2})} B_{12},
\end{equation}
where, in the last equality, we have defined the Bayes factor, $B_{12}$, as the ratio of model evidences. The Bayes factor reduces to the ratio of posterior probabilities in the case of equal prior probabilities $p(M_{1})$, $p(M_{2})$, which is usually assumed. The value of the Bayes factor can be interpreted on the empirical Jeffreys' scale \cite{Jeffreys:1961, Kass:1995} in order to determine the preference for one model over another. 

Moving to model comparison, we compute the evidence ratio for curved CDM and $\Lambda$CDM and find $\log B_{12} \approx -4.7$\footnote{Here, $\log$ denotes the logarithm to base $e$.}. Interpreted on the revised Jeffreys' scale \cite{Kass:1995}, this value denotes decisive evidence against curved CDM, which is in contrast to the results obtained for model rejection. 

This suggests that the low significance with which SNe Ia data exclude curved CDM is a consequence of assessing the goodness of fit of a given model, i.e. performing model rejection, rather than determining the preference of one model over another, i.e. performing model comparison. 

\section{Conclusions}\label{sec:conclusions}

In this work, we have investigated the use of the relative entropy to perform consistency tests in cosmology.

In a first part, we have revisited the relative entropy as a consistency measure between constraints obtained from different data sets, focusing on some of its key properties, such as asymmetry and path dependence. Taking into account both these properties, we have discussed how the Surprise statistic \cite{Seehars:2014} can be applied to assessing consistency between data sets for which the choice of prior and posterior in the computation of the relative entropy is ambiguous. Finally, we have illustrated these concepts in a simple toy model.

In a second part, we have proposed a novel model rejection method based on combining the KL divergence and the posterior predictive distribution. In our algorithm, we assess consistency between data and model by computing the relative entropy between the PPD derived from the data alone and that derived from data and model. This allows us to assess the goodness of fit of any given model, without considering an alternative. Some of the advantages of this method are that the PPDs are quite close to Gaussian in most cases, it is reasonably fast to implement and inconsistencies can be easily quantified with a $p$-value. We have demonstrated the method in a series of toy models. In order to test the applicability of the method to cosmological data, we have further applied it to SNe Ia data from the JLA \cite{Betoule:2014} and CMB constraints from Ref.~\cite{Planck-Collaboration:2016ai} in form of the shift parameter $R$. We have tested six cosmological models, i.e. (i) CDM, (ii) curved CDM, (iii) $\Lambda$CDM, (iv) curved $\Lambda$CDM, (v) $w_{0}$CDM and (vi) $w_{0}w_{a}$CDM and we found that SNe Ia data alone rule out CDM with high significance, but they cannot unambiguously distinguish between curved CDM and accelerated models. Investigating this further, we have performed a similar analysis using the Bayesian evidence instead of the relative entropy, finding comparable results. This suggests that the low significance with which SNe Ia data exclude a curved CDM cosmological model is a generic result of any model rejection analysis. When we add the CMB constraint to the JLA data, we find that both the CDM and the curved CDM model are clearly ruled out, whereas all other models provide a good fit to the data. Particularly, we find that $\Lambda$CDM is allowed by the combination of SNe Ia and CMB data. These results are consistent with previous works (e.g. \cite{Rubin:2016, Haridasu:2017, Nielsen:2016, Matthews:2017}), thus demonstrating that model rejection based on relative entropy is applicable to cosmological data and provides a promising method for upcoming analyses. In future work, we aim to explore this method further by investigating if it can be extended to model comparison and applying it to additional data.

\acknowledgments

\noindent We would like to thank Eiichiro Komatsu for many helpful comments on a previous version of this manuscript. We further thank the authors of \texttt{MCEvidence} for making their code public and the anonymous referee for helpful comments and suggestions. AN would also like to thank the organizers of the nonlinear Universe conference for the opportunity to present parts of this work. 

\noindent Based on observations obtained with Planck (http://www.esa.int/Planck), an ESA science mission with instruments and contributions directly funded by ESA Member States, NASA, and Canada.

\noindent The colors employed in this work are taken from $\tt{http://colorpalettes.net}$.

\clearpage

\appendix

\section{Path dependence of the relative entropy}\label{ap:path_dependence}

As discussed in Sec.~\ref{sec:sequential_updating}, the relative entropy is not additive for sequential updates, i.e. in the case illustrated in Fig.~\ref{fig:diagram}, we have
\begin{eqnarray}
D_{\mathrm{KL}, 0} \equiv D_{\mathrm{KL}}(p_{12}(\boldsymbol{\theta})||p(\boldsymbol{\theta})) &\neq D_{\mathrm{KL}}(p_{1}(\boldsymbol{\theta})||p(\boldsymbol{\theta})) + D_{\mathrm{KL}}(p_{12}(\boldsymbol{\theta})||p_{1}(\boldsymbol{\theta})), \\
D_{\mathrm{KL}, 0} \equiv D_{\mathrm{KL}}(p_{12}(\boldsymbol{\theta})||p(\boldsymbol{\theta})) &\neq D_{\mathrm{KL}}(p_{2}(\boldsymbol{\theta})||p(\boldsymbol{\theta})) + D_{\mathrm{KL}}(p_{12}(\boldsymbol{\theta})||p_{2}(\boldsymbol{\theta})).
\end{eqnarray}

The expected relative entropy, defined in Eq.~\ref{eq:exp_rel_ent}, does not depend on the specific experimental outcome $\textbf{y}_{i}$ by definition,
and can be rewritten using Eq.~\ref{eq:dkl_pr1} and Bayes' Theorem to obtain
\begin{equation}
\langle D_{\mathrm{KL}}(p_{1}(\boldsymbol{\theta})||p(\boldsymbol{\theta})) \rangle = \iint\mathrm{d}\textbf{y}_{1}\mathrm{d}\boldsymbol{\theta} \: p(\boldsymbol{\theta}, \textbf{y}_{1}) \log \frac{p(\boldsymbol{\theta}, \textbf{y}_{1})}{p(\boldsymbol{\theta})p(\textbf{y}_{1})}.
\end{equation}
This is known as the mutual information $\mathscr{I}(\textbf{y}_{1}; \boldsymbol{\theta})$ between $\textbf{y}_{1}$ and $\boldsymbol{\theta}$ and is further equivalent to Lindley's information measure for a given experiment \cite{Lindley:1956}. The mutual information quantifies the reduction of uncertainty in $\textbf{y}_{1}$ due to the knowledge of $\boldsymbol{\theta}$ \cite{Cover:2006}. It is symmetric, i.e. $\mathscr{I}(\textbf{y}_{1}; \boldsymbol{\theta}) = \mathscr{I}(\boldsymbol{\theta}; \textbf{y}_{1})$, and satisfies a chain rule for a sequence of experiments with outcomes $\textbf{y}_{i}$ \cite{Cover:2006}
\begin{equation}
\mathscr{I}(\textbf{y}_{1}, \textbf{y}_{2}, \cdots, \textbf{y}_{n}; \boldsymbol{\theta}) = \sum_{i} \mathscr{I}(\textbf{y}_{i}; \boldsymbol{\theta} \vert \textbf{y}_{i-1}, \textbf{y}_{i-2}, \cdots, \textbf{y}_{1}).
\label{eq:chain_rule}
\end{equation}
Here $\mathscr{I}(\textbf{y}_{i}; \boldsymbol{\theta} \vert \textbf{y}_{i-1}, \textbf{y}_{i-2}, \cdots, \textbf{y}_{1})$ denotes the mutual information between $\textbf{y}_{i}$ and $\boldsymbol{\theta}$ conditional on $\textbf{y}_{i-1}, \textbf{y}_{i-2}, \cdots, \textbf{y}_{1}$. The case illustrated in Fig.~\ref{fig:diagram} corresponds to $n = 2$ and  thus Eq.~\ref{eq:chain_rule} reduces to
\begin{equation}
\mathscr{I}(\textbf{y}_{1}, \textbf{y}_{2}; \boldsymbol{\theta}) = \mathscr{I}(\textbf{y}_{1}; \boldsymbol{\theta}) + \mathscr{I}(\textbf{y}_{2}; \boldsymbol{\theta} \vert \textbf{y}_{1}).
\end{equation}
Applying the chain rule for mutual information to the expected relative entropy, we obtain:
\begin{equation}
\langle D_{\mathrm{KL}}(p_{12}(\boldsymbol{\theta})||p(\boldsymbol{\theta})) \rangle = \langle D_{\mathrm{KL}}(p_{1}(\boldsymbol{\theta})||p(\boldsymbol{\theta})) \rangle + \langle D_{\mathrm{KL}}(p_{12}(\boldsymbol{\theta})||p_{1}(\boldsymbol{\theta})) \rangle,
\end{equation}
where we have defined\footnote{We note that this definition is an extension of the expected relative entropy given in Ref.~\cite{Seehars:2014}, as here we average over realizations of both $\textbf{y}_{1}$ and $\textbf{y}_{2}$.}
\begin{equation}
\langle D_{\mathrm{KL}}(p_{12}(\boldsymbol{\theta})||p_{1}(\boldsymbol{\theta})) \rangle = \iiint\mathrm{d}\boldsymbol{\theta}\mathrm{d}\textbf{y}_{1}\mathrm{d}\textbf{y}_{2} \: p(\boldsymbol{\theta}, \textbf{y}_{1}, \textbf{y}_{2}) \log \frac{p(\boldsymbol{\theta}, \textbf{y}_{2} \vert \textbf{y}_{1})}{p(\textbf{y}_{2} \vert \textbf{y}_{1})p_{1}(\boldsymbol{\theta})}.
\end{equation}
If we equivalently consider $\textbf{y}_{2}$ as the first and $\textbf{y}_{1}$ as the second experiment, we obtain
\begin{equation}
\langle D_{\mathrm{KL}}(p_{12}(\boldsymbol{\theta})||p(\boldsymbol{\theta})) \rangle = \langle D_{\mathrm{KL}}(p_{2}(\boldsymbol{\theta})||p(\boldsymbol{\theta})) \rangle + \langle D_{\mathrm{KL}}(p_{12}(\boldsymbol{\theta})||p_{2}(\boldsymbol{\theta})) \rangle.
\end{equation}
This means that the expected information gain from a prior to a posterior is additive for sequential updates, as opposed to the observed relative entropy.

\section{Assessing consistency for arbitrary updates}\label{ap:dkl_asymmetry}

In Sec.~\ref{sec:rel_ent_consistency}, we have argued that the path dependence of the relative entropy can cause data set inconsistencies to only be detectable in one of the two updates defined in Equations \ref{eq:dkl_1comb} and \ref{eq:dkl_2comb}. We can alternatively understand this case through the Shannon and the cross-entropy. To this end, we can rewrite Equations \ref{eq:dkl_1comb} and \ref{eq:dkl_2comb} as
\begin{eqnarray}
D_{\mathrm{KL}, 3} = H(p_{12}(\boldsymbol{\theta}), p_{1}(\boldsymbol{\theta})) - H(p_{12}(\boldsymbol{\theta})),\\
D_{\mathrm{KL}, 4} = H(p_{12}(\boldsymbol{\theta}), p_{2}(\boldsymbol{\theta})) - H(p_{12}(\boldsymbol{\theta})).
\end{eqnarray}
The quantity $H(q(\boldsymbol{\theta}))$ denotes the generalization of the Shannon entropy for continuous random variables, also called differential entropy, which for a pdf $q$, is defined as \cite{Cover:2006}
\begin{equation}
H(q(\boldsymbol{\theta})) = - \int \mathrm{d}\boldsymbol{\theta} \; q(\boldsymbol{\theta}) \log q(\boldsymbol{\theta}).
\end{equation} 
Furthermore, $H(q_{1}(\boldsymbol{\theta}), q_{2}(\boldsymbol{\theta}))$ is the cross-entropy between distributions $q_{1}(\boldsymbol{\theta})$ and $q_{2}(\boldsymbol{\theta})$, i.e. \cite{Good:1956, Kerridge:1961}
\begin{equation}
H(q_{1}(\boldsymbol{\theta}), q_{2}(\boldsymbol{\theta})) = - \int \mathrm{d}\boldsymbol{\theta} \; q_{1}(\boldsymbol{\theta}) \log q_{2}(\boldsymbol{\theta}).
\end{equation} 
The cross-entropy is a similarity measure between a pdf $q_{1}(\boldsymbol{\theta})$ and its approximation $q_{2}(\boldsymbol{\theta})$ and is minimized when $q_{1}(\boldsymbol{\theta}) = q_{2}(\boldsymbol{\theta})$ \cite{Good:1956, Kerridge:1961}. 

A relative entropy between the combined distribution and one of the single-experiment distributions that is significantly larger than expected thus suggests a large cross-entropy between the two. This means that the two distributions are significantly different. Once again we therefore see that it is advisable to reject the null hypothesis of consistency if either one of the relative entropies in Equations \ref{eq:dkl_1comb} and \ref{eq:dkl_2comb} is significantly larger than expected a priori. 

\section{Gaussianity tests}\label{ap:gaussianity_tests}

\begin{figure}
\begin{center}
\includegraphics[scale=0.45]{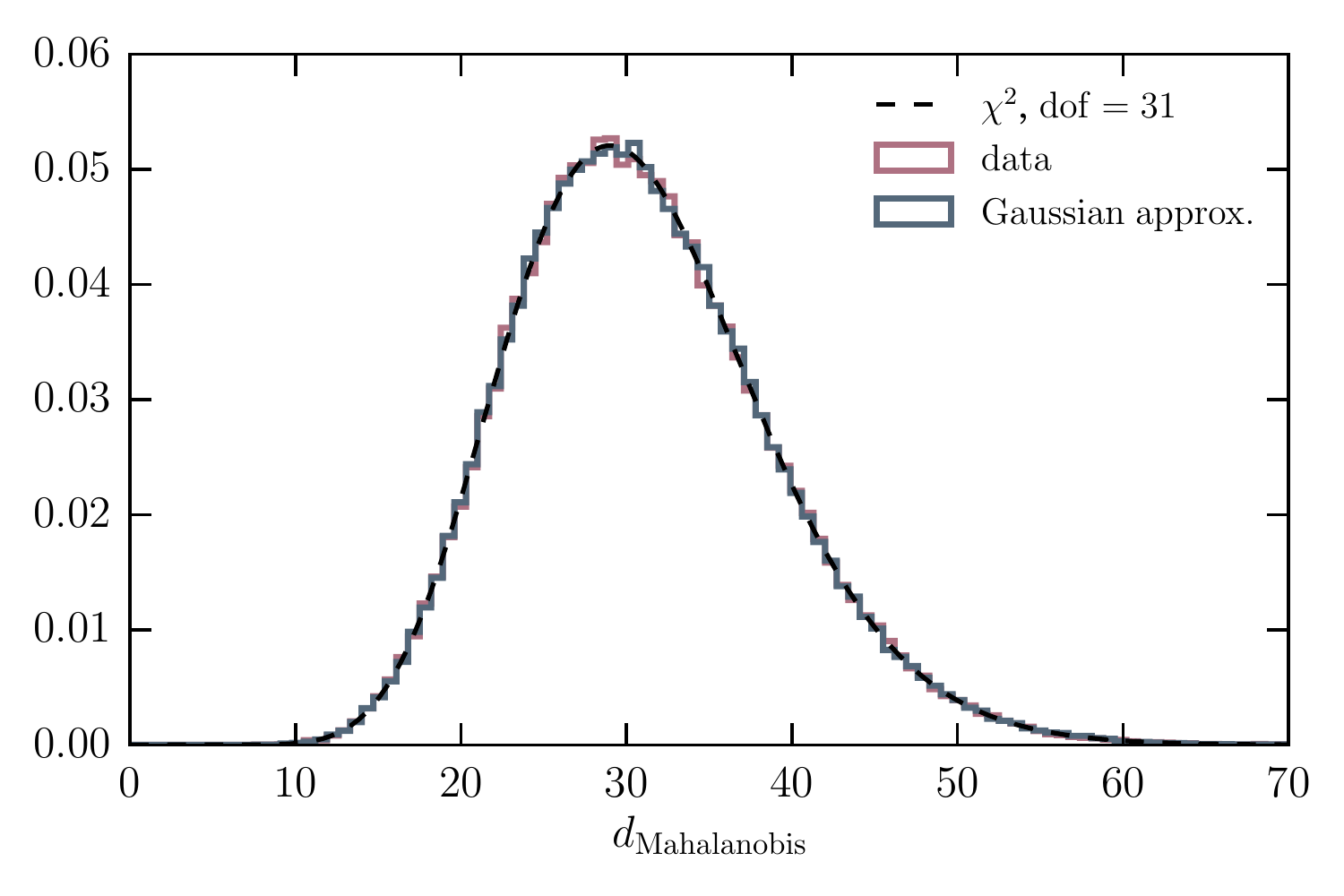}
\caption{Distribution of Mahalanobis distances for the PPD derived from data and model and its Gaussian approximation using data from SNe and CMB in the framework of curved CDM. We also show the theoretically expected $\chi^{2}$ distribution for 31 degrees of freedom. For this plot, we have set the number of realizations to $N=200$.}
\label{fig:data+model_post_dMahalanobis}
\end{center}
\end{figure}

In order to compute relative entropies using the analytic expressions derived in Ref.~\cite{Seehars:2014}, we need to test that the relevant distributions are well-approximated by Gaussians. In this work, we perform two different tests: (i) we visually check that the Mahalanobis distances \cite{Mahalanobis:1936} of the distributions follow a $\chi^{2}$ distribution with number of degrees of freedom equal to the data vector dimension and (ii) we compare the marginalized PPD constraints to their Gaussian approximations. Both these tests show good agreement with Gaussianity. In Fig.~\ref{fig:data+model_post_dMahalanobis} we show the distribution of the Mahalanobis distances for the PPD derived from data and model and its Gaussian approximation for SNe and CMB data in curved CDM, as an example. From the figure, we see that the Mahalanobis distances of the PPD indeed follow a $\chi^{2}$ distribution with 31 of degrees of freedom, as expected for Gaussian data\footnote{We do not show the marginalized PPD contours as the parameter space is 30-dimensional but a visual inspection also shows very good agreement between the original and the Gaussian samples.}. The results for all other distributions are similar and we thus do not show them here.

\section{Alternative method for computing the relative entropy}\label{ap:alternat_rel_ent_alg}

As discussed in Sec.~\ref{subsec:implementation}, there exists an alternative method for computing the relative entropies in the case in which we can directly sample from the PPDs. We illustrate the method for the computation of $D_{\mathrm{KL}}(\textbf{y}, M || \textbf{y})$, but all other quantities can be obtained similarly. In a first step we directly create a sample of length $N \times \bar{N}$ from each of the PPDs $p(\textbf{y}^{\prime} \vert \textbf{y}, C_{\textbf{y}}, C_{\textbf{y}^{\prime}}, M)$ and $p(\textbf{y}^{\prime} \vert \textbf{y}, C_{\textbf{y}}, C_{\textbf{y}^{\prime}})$. We then compute the relative entropy $D_{\mathrm{KL}}(\textbf{y}, M || \textbf{y})$ between prior and posterior from these samples. This method is significantly faster than the one described in Sec.~\ref{subsec:implementation} and is thus preferable in cases in which direct sampling from the PPDs is possible.

\section{$p$-value computation through realized discrepancies}\label{ap:realized_disc}

Ref.~\cite{Gelman:1996} proposed an algorithm to compute posterior predictive $p$-values based on comparing so-called discrepancy statistics between the observed data and model predictions to those obtained comparing the model to simulated data. Applying it to our case, we obtain (c.f. Sec.~\ref{subsec:cosmology_applic}):
\begin{enumerate}
\item Draw a realization $\boldsymbol{\theta}_{i}$ of model parameters from the posterior $p(\boldsymbol{\theta} \vert \textbf{y}, C_{\textbf{y}})$.
\item Compute a corresponding sample of the data, $\textbf{y}_{\mathrm{sim}}(\boldsymbol{\theta}_{i})$.
\item Compute $D_{\mathrm{KL}}(\textbf{y}, M || \textbf{y})$ and $D_{\mathrm{KL}}(\textbf{y}, M || \textbf{y}_{\mathrm{sim}}(\boldsymbol{\theta}_{i}))$.
\end{enumerate}
Finally, we calculate the $p$-value of the observed $D_{\mathrm{KL}}(\textbf{y}, M || \textbf{y})$ as the fraction of draws for which $D_{\mathrm{KL}}(\textbf{y}, M || \textbf{y}_{\mathrm{sim}}(\boldsymbol{\theta}_{i})) > D_{\mathrm{KL}}(\textbf{y}, M || \textbf{y})$.

We apply this alternative algorithm to compute the $p$-values of the observed relative entropies when fitting the cosmological models given in Tab.~\ref{tab:models} to SNe Ia data alone. As in Sec.~\ref{subsec:implementation}, we bin the SNe Ia data into 30 equally log-spaced bins in the redshift range $0.01 < z < 1.3$. We find results very similar to those shown in Sec.~\ref{subsec:implementation} and we therefore do not show them here. This analysis shows the stability of the method chosen to compute the $p$-values and the applicability of the realized discrepancy algorithm to our formalism.

\section{Toy model implementation details}\label{ap:toy_model_implementation}

\subsection{Implementation choices}

In the toy model described in Sec.~\ref{subsec:toy_model_model_selection}, we consider fitting data to polynomials of varying degree $\hat{n}$. For each polynomial of degree $n$, we assume the underlying true model to be given by 
\begin{equation}
y_{i} = \sum_{k=0}^{n} c_{k} x_{i}^{k}.
\end{equation}
The data $y_{\mathrm{d}, i}$ is further assumed to be normally distributed around the true values $y_{i}$ with a constant standard deviation $\sigma$, i.e.
\begin{equation}
y_{\mathrm{d}, i} = y_{i} + \epsilon_{i}, \; i=1, \ldots, m,
\end{equation}  
where $\epsilon_{i}$ is drawn from a Gaussian with mean zero and standard deviation $\sigma$, i.e. $\mathcal{N}(0, \sigma)$. In the specific implementation described in Sec.~\ref{subsec:toy_model_model_selection}, we choose the dimension of the data vector as $m = 10$, the $x$-values, $x_{i}$, linearly spaced in $[0.1, \; 2.2]$ and the standard deviation of the data as $\sigma = 0.2$.

\subsection{Choice of fiducial polynomial coefficients}

We choose the fiducial polynomial coefficients in a similar way as we would for real data: we first determine fiducial coefficients for an 8-degree polynomial, which we choose to be given by $\textbf{c} = \big(3., \allowbreak 1.5, \allowbreak 1., \allowbreak 0.5, \allowbreak 1.4, \allowbreak 0.1, \allowbreak 0.7, \allowbreak 0.3, \allowbreak 3.\big)$\footnote{These coefficients are not completely random, as we need to make sure that the different fiducial models result in sufficiently different data compared to the uncertainties such that we have the statistical power to distinguish the considered models.}. We then generate a realization $\textbf{y}_{\mathrm{d}}$ thereof with covariance matrix $C_{\textbf{y}}$. This represents the observed data. We then set the fiducial polynomial coefficients for each degree $n = 0, \ldots, 7$ considered in our analysis to the best-fit coefficients for this particular model and $\textbf{y}_{\mathrm{d}}$. This procedure ensures that the choice of fiducial polynomial in our analysis is driven by the data, which is the case in any application of this method to real data. 

\section{SNe Ia and CMB posterior means}\label{ap:posteriors}

In this section, we give the posterior means obtained for the cosmological analyses performed in this work: we show the constraints obtained for the analysis  of JLA SNe Ia data alone in Tab.~\ref{tab:constraints_SNe}. The constraints for the combined analysis of JLA data and the CMB shift parameter $R$ are shown in Tab.~\ref{tab:constraints_SNe+CMB}. Where we can compare, our results agree well with those given in Ref.~\cite{Betoule:2014}.

\begin{landscape}
\begin{table*}
\caption{Summary of SNe Ia posterior means for all cosmological models considered in this work.} \label{tab:constraints_SNe}
\begin{center}
\begin{tabular}{cccccccccc}
\hline\hline 
Model  & $\Omega_{\mathrm{m}}$ & $\Omega_{\Lambda}$ & $w_{0}$ & $w_{a}$ & $\alpha$ & $\beta$ & $M^{1}_{\mathrm{B}}$ & $\Delta M$ \\ \hline     
$\Lambda$CDM & $0.297 \pm 0.034$ & - & - & - & $0.141 \substack{+0.007 \\ -0.006}$ & $3.10 \pm 0.08$ & $-19.05 \pm 0.02$ & $-0.070 \pm 0.023$ \\
CDM & - & - & - & - & $0.137 \pm0.007$ & $3.16 \pm 0.08$ & $-18.80 \pm 0.02$ & $-0.087 \pm 0.023$ \\
c. CDM  & $0.030 \substack{+0.017 \\ -0.026}$ & - & - & - & $0.139 \pm 0.007$ & $3.09 \pm 0.08$ & $-19.00 \pm 0.02$ & $-0.074 \pm 0.023$ \\
c. $\Lambda$CDM  & $0.206 \substack{+0.107 \\ -0.102}$ & $0.563 \substack{+0.154 \\ -0.156}$ & - & - & $0.141 \pm 0.007$ & $3.11 \pm 0.08$ & $-19.04 \pm 0.03$ & $-0.070 \pm 0.023$ \\
$w_{0}$CDM & $0.254 \substack{+0.092 \\ -0.097}$ & - & $-0.93 \pm 0.20$ & - & $0.141 \pm 0.007$ & $3.10 \pm 0.08$ & $-19.04 \pm 0.03$ & $-0.070 \pm 0.023$ \\
$w_{0}w_{a}$CDM & $0.280 \substack{+0.095 \\ -0.100}$ & - & $-0.95 \pm 0.22$ & $-0.30 \substack{+0.98 \\ -1.03}$ & $0.141 \pm 0.007$ & $3.10 \pm 0.08$ & $-19.04 \pm 0.03$ & $-0.071 \pm 0.023$ \\
 \hline \hline
\end{tabular}
\end{center}
\end{table*}

\begin{table*}
\caption{Summary of SNe Ia and CMB posterior means for all cosmological models considered in this work.} \label{tab:constraints_SNe+CMB}
\begin{center}
\begin{tabular}{ccccccc}
\hline\hline 
Model & $h$ & $\Omega_{\mathrm{m}}$ & $\Omega_{\mathrm{b}}$ & $\Omega_{\Lambda}$ & $w_{0}$ & $w_{a}$ \\ \hline     
$\Lambda$CDM & $0.75 \substack{+0.31 \\ -0.30}$ & $0.297 \substack{+0.017 \\ -0.018}$ & $0.050 \pm 0.027$ & - & - & -  \\
CDM & $0.202 \pm 0.002$ & - & $0.075 \substack{+0.012 \\ -0.013}$ & - & - & -  \\
c. CDM & $0.202 \pm 0.002$ & $0.999 \pm 0.001$ & $0.074 \substack{+0.012 \\ -0.013}$ & - & - & - \\
c. $\Lambda$CDM & $0.69 \substack{+0.35 \\ -0.34}$ & $0.297 \pm 0.037$ & $0.050 \substack{+0.027 \\ -0.028}$ & $0.701 \pm 0.028$ & - & -  \\
$w_{0}$CDM & $0.71 \substack{+0.33 \\ -0.34}$ & $0.303 \pm 0.020$ & $0.050 \pm 0.027$ & - & $-1.00 \substack{+0.07 \\ -0.06}$ & - \\
$w_{0}w_{a}$CDM & $0.69 \pm 0.34$ & $0.298 \pm 0.026$ & $0.050 \substack{+0.028 \\ -0.027}$ & - & $-0.91 \pm 0.17$ & $-0.57 \pm 0.87$ \\
 \hline \hline
\end{tabular}
\end{center}
\end{table*}

\addtocounter{table}{-1}

\begin{table*}
\caption{Summary of SNe Ia and CMB posterior means for all cosmological models considered in this work (\textit{continuation}).}
\begin{center}
\begin{tabular}{ccccccccccc}
\hline\hline 
Model & $\alpha$ & $\beta$ & $M^{1}_{\mathrm{B}}$ & $\Delta M$ \\ \hline     
$\Lambda$CDM & $0.141 \pm 0.007 $ & $3.10 \pm 0.08$ & $-19.05 \pm 0.02$ & $-0.070 \pm 0.023$ \\
CDM & $0.137 \pm 0.007 $ & $3.15 \pm 0.08$ & $-18.80 \pm 0.02$ & $-0.087 \pm 0.022$ \\
c. CDM & $0.136 \pm 0.007 $ & $3.16 \pm 0.08$ & $-18.80 \pm 0.02$ & $-0.086 \pm 0.022$ \\
c. $\Lambda$CDM & $0.141 \pm 0.007 $ & $3.11 \pm 0.08$ & $-19.05 \pm 0.02$ & $-0.070 \pm 0.022$ \\
$w_{0}$CDM & $0.141 \pm 0.007 $ & $3.11 \pm 0.08$ & $-19.05 \pm 0.02$ & $-0.070 \substack{+0.022 \\ -0.023}$ \\
$w_{0}w_{a}$CDM & $0.141 \pm 0.007 $ & $3.10 \pm 0.08$ & $-19.04 \pm 0.03$ & $-0.070 \pm 0.023$ \\
 \hline \hline
\end{tabular}
\end{center}
\end{table*}
\end{landscape}

\section{Error model tests}\label{ap:test-sampling}

\begin{figure}
\begin{center}
\subfigure{\includegraphics[width=0.49\textwidth]{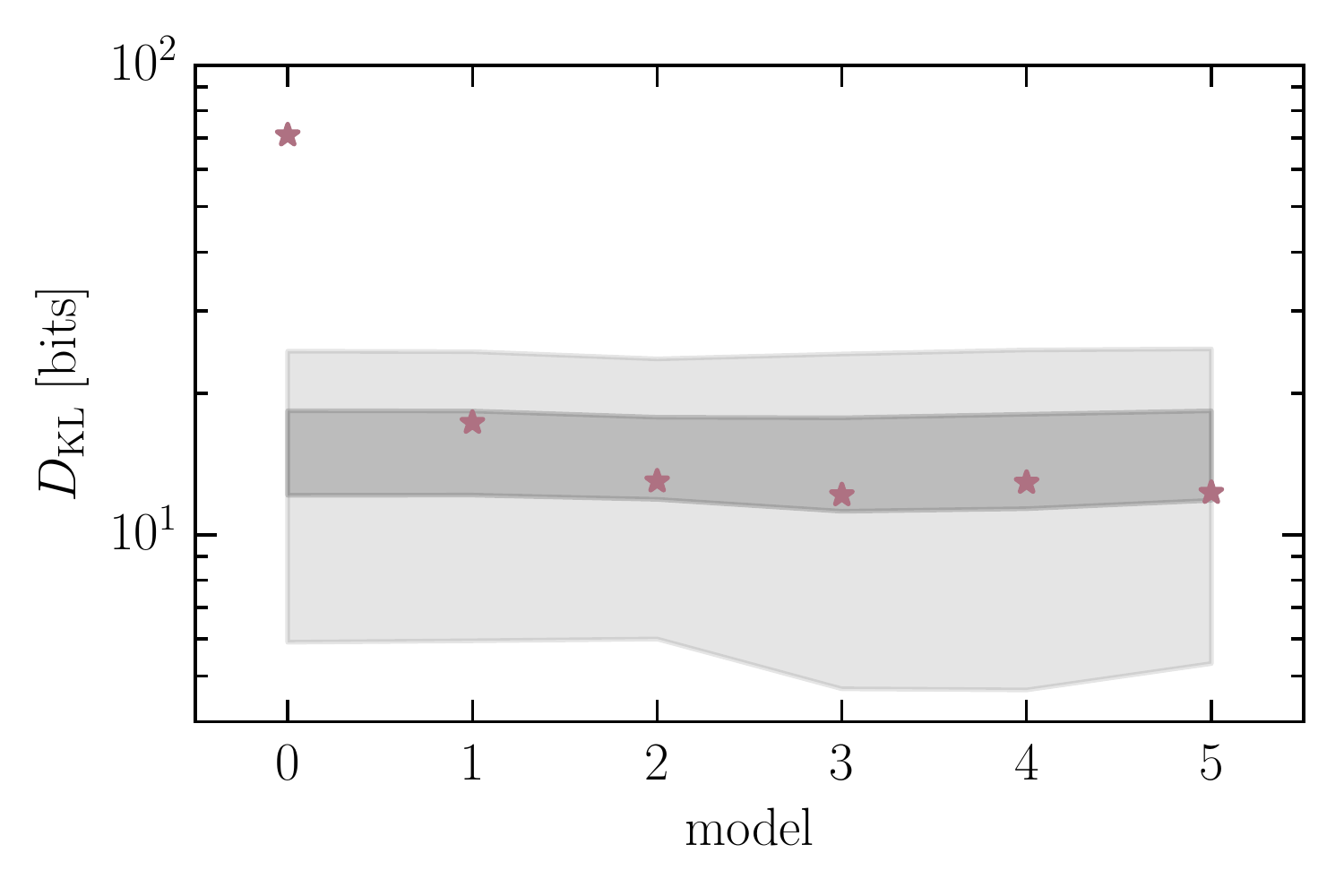}}
\subfigure{\includegraphics[width=0.49\textwidth]{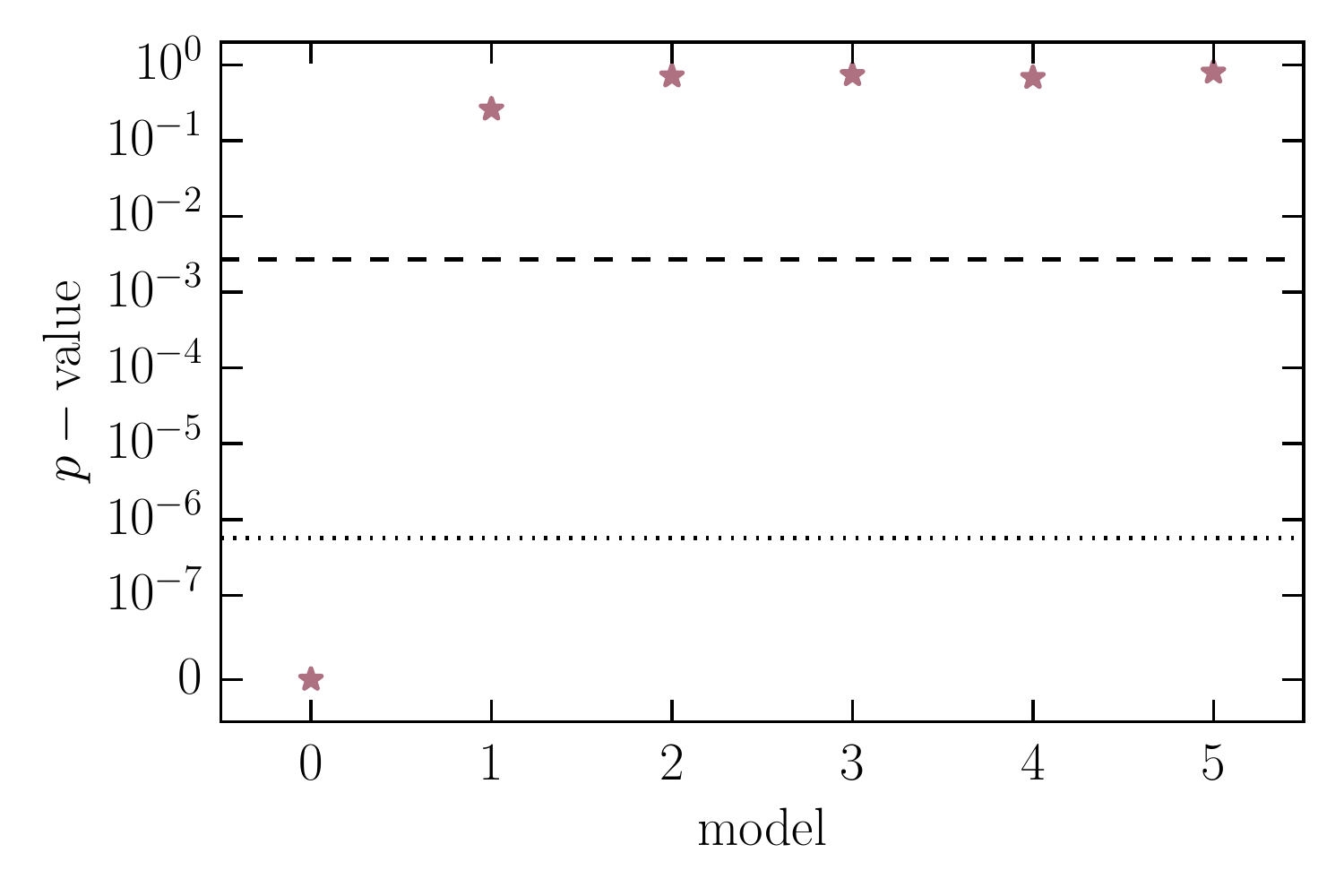}} 
\caption{Illustration of $D_{\mathrm{KL}}(\boldsymbol{\mu}, M_{i} || \boldsymbol{\mu})$ and the $p$-value of $D_{\mathrm{KL}}(\boldsymbol{\mu}, M_{i} || \boldsymbol{\mu})$ obtained for a fit to JLA SNe Ia as a function of the cosmological model chosen to fit the data. The expected relative entropies are computed using Particle Swarm Optimization. The gray bands show the $1\sigma$ and $3\sigma$ errors on the information gain from the true model, i.e. $\langle D_{\mathrm{KL}}(\boldsymbol{\mu}(M_{i}), M_{i} || \boldsymbol{\mu}(M_{i})) \rangle$. The dashed, horizontal line denotes a $p$-value of $2.7 \times 10^{-3}$, while the dotted line illustrates a $p$-value of $5.7 \times 10^{-7}$. These correspond to the Gaussian equivalent $3 \sigma$ and $5 \sigma$ thresholds. The correspondence between model number and cosmological model is given in Tab.~\ref{tab:models}.} 
\label{fig:dkl_vs_model_jla_test_implementation}
\end{center}
\end{figure}

As discussed in Sec.~\ref{subsec:cosmology_applic}, it is prohibitively expensive to determine the best-fit model parameters for each simulated data set when computing the expected relative entropy. In order to avoid this step, we first sample a realization of model parameters and then determine a corresponding data realization by sampling from the distribution of residual errors. We test this method by estimating the expected relative entropy in an alternative way: for each simulated data realization, we determine the best-fit model parameters using a Particle Swarm Optimizer (PSO). We further assume that the model parameter covariance does not significantly depend on the parameter values. Therefore, we keep the covariance matrix constant for each iteration and set it to the model parameter covariance determined at the best-fit parameters to the real data. Finally, we compute the relative entropy between the ppd from data and the ppd from data and model for each pair of simulated data and corresponding best-fit model parameters. We apply this method to the JLA SNe Ia data and Fig.~\ref{fig:dkl_vs_model_jla_test_implementation} shows the relative entropies and associated $p$-values obtained for each cosmological model considered in our analysis. These results are consistent with those obtained using the approximate method, as can be seen from comparing with Fig.~\ref{fig:dkl_vs_model_jla}. We therefore conclude that the simplified method results in an acceptable approximation to the full method and we therefore resort to the former in the rest of this work.

\bibliography{main_text_incl_figs}

\end{document}